\documentclass{aa}
\usepackage{txfonts}
\usepackage{graphicx}
\usepackage{longtable}
\usepackage{lscape}

\newcommand{\e}{et al.\ }
\begin{document}

\title{
The stellar population of Sco OB2 revealed by Gaia DR2 data
\thanks{Tables~\ref{table-pms} and~\ref{table-upp-ms} are only available in
electronic form at the CDS via anonymous ftp to cdsarc.u-strasbg.fr
(130.79.125.5) or via http://cdsweb.u-strasbg.fr/Abstract.html}
}

\date{Received date / Accepted date}

\author{F. Damiani\inst{1}
\and L. Prisinzano\inst{1} \and I. Pillitteri\inst{1}
\and G. Micela\inst{1} \and S. Sciortino\inst{1}
}
\institute{INAF - Osservatorio Astronomico di Palermo G.S.Vaiana,
Piazza del Parlamento 1, I-90134 Palermo, ITALY \\
\email{francesco.damiani@inaf.it}
}

\abstract{
The Sco~OB2 association is the nearest OB association, extending over
approximately 2000 square degrees on the sky. Only its brightest and
most massive members are already known (from Hipparcos) across its
entire size, while studies of its lower-mass population refer only to
small portions of its extent.
}{
In this work we exploit the capabilities of Gaia DR2 measurements to
search for Sco~OB2 members across its entire size and down to the lowest
stellar masses.
}{
We use both Gaia astrometric (proper motions and parallaxes) and photometric
measurements (integrated photometry and colors) to select association
members, using minimal assumptions derived mostly from the Hipparcos
studies. Gaia resolves small details in both the kinematics of
individual Sco~OB2 subgroups and their distribution with distance from
the Sun.
{
Methods are developed to explore the three-dimensional kinematics of a
stellar population covering large sky areas.
}
}{
We find { nearly 11\, 000 pre-main sequence
members (with less than 3\% field-star contamination)
of Sco~OB2, plus $\sim 3600$ main-sequence candidate members with a larger
(10-30\%) field-star contamination.
A higher-confidence subsample of $\sim 9200$ pre-main-sequence (and
$\sim 1340$ main-sequence) members is also selected ($<1$\%
contamination for the pre-main-sequence),
affected however by larger ($\sim 15$\%) incompleteness.}
{ We classify separately stars in compact and diffuse populations.
Most members belong to one of several kinematically distinct diffuse
populations, whose ensemble}
outlines clearly the shape of the entire association. Upper Sco is
the densest region of Sco OB2. It is characterized by a complex spatial and
kinematical structure, with no global pattern of motion. Other dense
subclusters are found in Upper Centaurus-Lupus (the richest one coincident
with the group
near V1062~Sco already found by R\"oser \e 2018), and in Lower Centaurus-Crux.
{ Most of} the clustered stars appear to be younger than the diffuse PMS
population, suggesting star formation in small groups which rapidly
disperse and dilute, reaching space densities lower than field stars
while keeping memory of their original kinematics.
We also find that the open cluster IC~2602 has a similar dynamics
to Sco~OB2, and its PMS members are currently evaporating and forming a
diffuse (size $\sim 10^\circ$) halo around its double-peaked core.
}{}


\titlerunning{The stellar population of Sco OB2 from Gaia DR2}
\authorrunning{Damiani et al.}

\maketitle

\section{Introduction}
\label{intro}

Sco~OB2 is the OB association nearest to the Sun. Its properties were
reviewed by Preibisch and Mamajek (2008), who summarize a very large
literature up to that date. Its close distance (of order of 120-160~pc),
together with the low spatial density of its probable members, makes a
comprehensive study of its total population very difficult. As
summarized by Preibisch and Mamajek (2008), there was a decade-long
debate on whether this assembly of B stars (only one O star,
$\zeta$~Oph, is known to belong to the association) represents one
physical entity or not. The very diluted appearance, and the similarity
between the proper motion of putative members with the reflex solar
motion were arguments against its real existence. The Hipparcos data
published by de~Zeeuw \e (1999) provided however evidence in favor of a
physical origin of the association, but were limited to its brightest
members. As remarked by Preibisch and Mamajek (2008), most of its
members are still undiscovered, if a normal IMF is present.
The Sco~OB2 association consists of three large sub-associations: Upper
Sco-Cen (USC), with still ongoing star formation in the $\rho$~Oph
dark clouds; Upper Cen-Lup (UCL) containing the Lupus dark clouds; and
Lower Cen-Crux (LCC), which crosses the Galactic plane from N to S, which is
thought to be the oldest part of Sco~OB2.

The huge apparent size of Sco~OB2 (approximately $80^\circ \times
30^\circ$), and its sky position towards the densest regions of the inner
Galactic plane, are big obstacles for a large-scale study of its
population across the entire stellar mass range. Existing large-scale
surveys such as PanSTARRS (Chambers \e 2016) or VPHAS+ (Drew \e 2014)
cover only a small fraction of the entire association.
In fact, the great majority of the literature on Sco~OB2 (too vast to be
reviewed here) consists of studies of spatial regions covering only a
small fraction of the association size.
Even X-ray imaging surveys, such as those by Krautter \e (1997) on Lupus or
Sciortino \e (1998) on Upper Sco yielded only 136 and 50 members,
respectively, over a combined sky region of $\sim 10$\% of the total Sco~OB2
extent.
{
Studies such as Pecaut \e (2012) or Pecaut and Mamajek (2016; PM16) on
the late-type Sco~OB2 population discovered less than 600 members,
selected using a large variety of methods; others like Hoogerwerf (2000)
are affected by substantial field-star contamination.
}
Instead, the Gaia observations (Gaia Collaboration \e 2016, 2018a)
are very well suited for a complete study, fulfilling the
requisites of homogeneity over the entire size of the association,
sufficient photometric depth to cover the entire mass range, and extremely high
astrometric precision to resolve ambiguities with reflex solar motion,
{
and to avoid strong field-star contamination.
}
Therefore, we devote this work to the study of the entire Sco~OB2
association using the Gaia DR2 data. Partial studies of the region have
already been presented by Wright and Mamajek (2018) using earlier Gaia DR1 data,
limited to the brighter stars in the Upper-Sco region. Using Gaia DR2
data, Manara \e (2018) have studied a small number of candidate members
in the Lupus clouds, of which only five were confirmed as members;
instead, Goldman \e (2018) report on the discovery of more than 1800
members in the Lower Cen-Crux part of Sco~OB2 down to substellar masses.

{
Throughout this work, we use exclusively Gaia data as far as possible,
even though there are many other datasets available on this association,
with a two-fold purpose: the first is maximum uniformity, since member samples
assembled in previous studies do not benefit from such a uniform
selection strategy as is now possible with Gaia; second, in doing this
we develop and test a method for studying young clusters and
associations using Gaia which might in principle be applied to a large
number of star-formation regions, for which less or no auxiliary data are
available to complement Gaia data.

An astrometric study of the census of a cluster or association has its
roots in the kinematical coherence of their members, inherited from the
bulk motion of their parent cloud. In the case of an association like
Sco~OB2, extended over tens of degrees in the sky, projection and depth effects
become important, and have actually been exploited in the studies based
on Hipparcos (convergent-point method), such as de~Zeeuw \e (1999) or de
Bruijne (1999). We discuss in depth the reconstruction of space
velocities based on the new Gaia data.
Since the Gaia DR2 parallaxes ($\pi$) used in this work are much
more precise than the Hipparcos parallaxes, some methods discussed here
were not directly applicable to the Hipparcos data.
It is also worth remarking that very young clusters are often not
``kinematically simple'', but show instead evidence of multiple
kinematical populations in the same sky region: some examples are found in
Jeffries \e (2014), Tobin \e (2015), Sacco \e (2015), or Damiani \e (2017).
}

This paper is structured as follows: Sect.~\ref{obs} describes the Gaia
data used.
{
Sect.~\ref{space} discusses analytical transformations between apparent
and space velocities that are relevant in this context.
}
Sect.~\ref{members} presents our criteria for member
selection. Sect.~\ref{spatial}
{
presents the selection of various Sco~OB2 subpopulations based on
spatial and kinematical criteria.
}
Sect.~\ref{ages}
discusses ages of { the subpopulations.
Sect.~\ref{three-dim} discuss the three-dimensional structure of the
association.
}
Finally, Sect.~\ref{concl} provides a concluding summary of results.

\section{Gaia Observations}
\label{obs}

\begin{figure}
\resizebox{\hsize}{!}{
\includegraphics[]{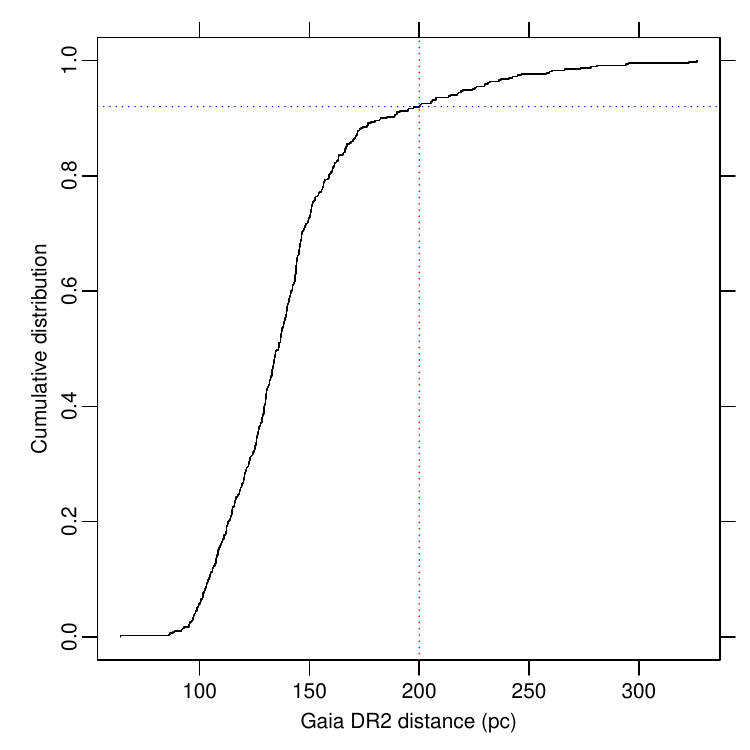}}
\caption{
{
Cumulative distribution of Gaia DR2 distances for Hipparcos
members of Sco~OB2. The vertical dotted line indicates our initial
survey limit (200~pc).
The horizontal dotted line indicates the percentage (92\%) of Hipparcos members
closer than this distance.
}
\label{dezeeuw-dist-cumfun}}
\end{figure}

With an estimated size of $\sim 2000$~sq.deg., and its location in the
inner Galactic plane, a blind search among all Gaia DR2 sources would
yield an unmanageable source list.
{
We therefore select Gaia sources up to some maximum distance. In
order to choose the distance limit, we matched the Hipparcos probable Sco~OB2
members from de Zeeuw \e (1999) with the Gaia DR2 catalog (keeping only
Gaia sources with small relative error on parallax, i.e.\ $\pi/\Delta \pi>10$),
and computed their cumulative distance distribution
(Fig.~\ref{dezeeuw-dist-cumfun}): this shows that while $\sim 90$\% of
these stars have distances $d$ in the range $100 < d < 170$~pc, a
curious tail including $\sim 8$\% of stars extends as far as $d \sim
300$~pc. In our search for Sco~OB2 members,
we decided to set an initial distance limit to 200~pc, and
examine later any evidence for association members between 200-300~pc.

In addition to the distance constraint, we note that
}
the large majority ($\sim 90$\%) of $\pi$ measurements
in Gaia DR2 are of low S/N ratio, and do not allow to locate stars with
great precision. Therefore, we also require that $\pi/\Delta \pi>10$
(column parallax\_over\_error from table gaiadr2.gaia\_source),
that is a relative
error on $\pi$ (and distance) less than 10\%, for a good three-dimensional
positioning of all candidate members.

\begin{figure}
\resizebox{\hsize}{!}{
\includegraphics[]{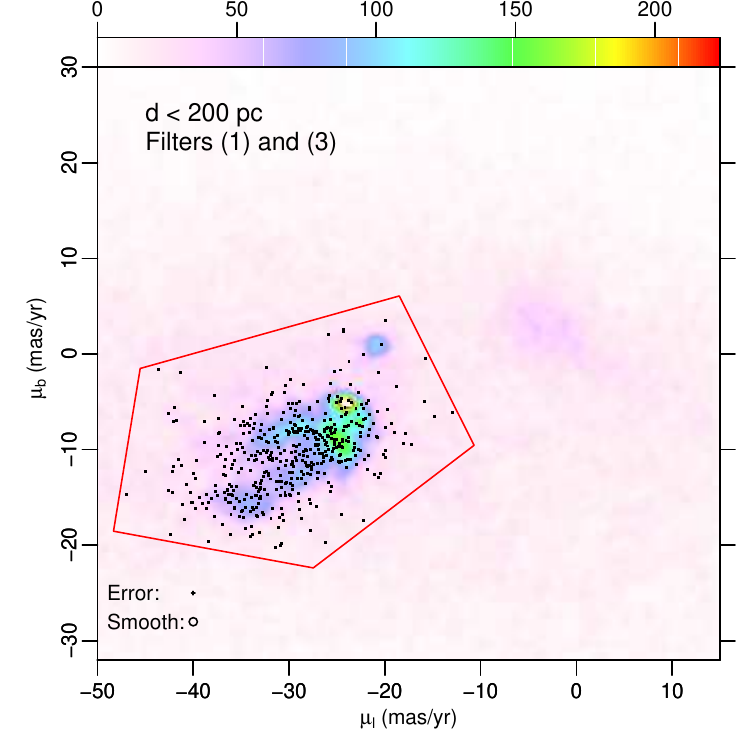}}
\caption{
Density plot (2-d histogram) of Gaia sources in the proper-motion plane
$(\mu_{l},\mu_{b})$ in Galactic coordinates, with parallax $\pi>5$ and
relative error on parallax $\pi/\Delta_{\pi}>10$, Gaussian-smoothed,
and with filters (1) and (3) applied (see Sect.\ref{obs} for details).
The circle at $(\mu_{l},\mu_{b}) = (-40,-28)$ indicates the $1-\sigma$ size
of the smoothing Gaussian.
The small cross at $(\mu_{l},\mu_{b}) = (-40,-25)$ represents the median
PM error.
Black dots are the Hipparcos Sco OB2 likely members from de Zeeuw \e (1999).
The color scale on the top axis indicates source density in
units of sources/(mas/yr)$^2$.
The red polygon enclosing all Hipparcos members { defines the}
extraction region of { PM-selected} Gaia Sco~OB2 { candidate} members.
\label{pm-select-init}}
\end{figure}

\begin{figure}
\resizebox{\hsize}{!}{
\includegraphics[]{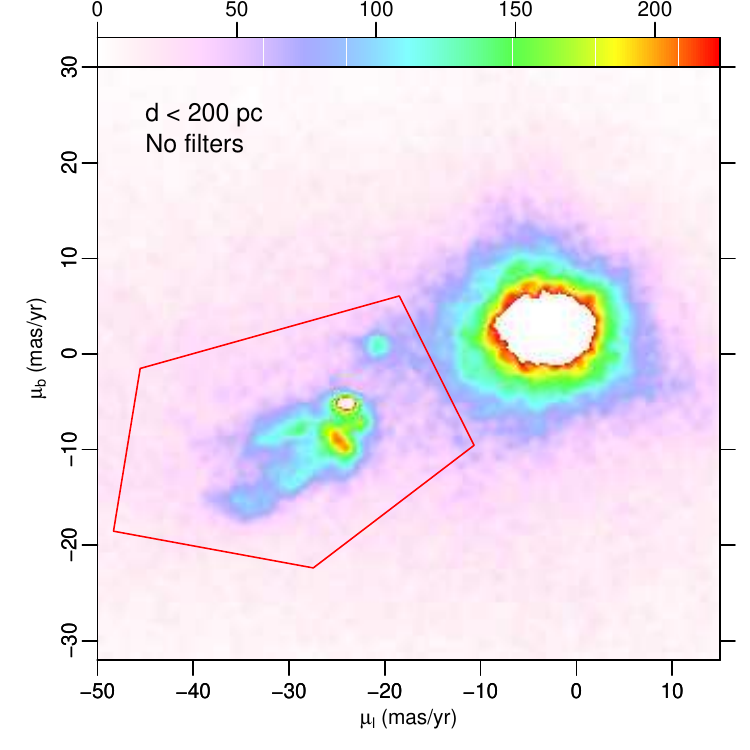}}
\caption{
Same as Fig.~\ref{pm-select-init}, but without filters (1) and (3).
The same color scale is employed:
densities above the range in the top-axis colorbar are displayed
in white (e.g.\ the core of the field-star distribution near (-3,0)).
\label{pm-nofilt}}
\end{figure}

The spatial region used for selection was chosen to be slightly larger than
that shown in de~Zeeuw \e (1999) and Preibisch and Mamajek (2008; their
Fig.~2), in order to determine the boundaries of Sco~OB2 in the most
unbiased way.
{
Therefore, we searched the Gaia DR2 catalog in the
galactic longitude range $280 <l< 360$ and galactic latitude range
$-10 <b< +30$.  }
This search, with the above constraints on $\pi>5$ and $\pi/\Delta \pi>10$,
{ resulted in 308\,260 Gaia DR2 sources.}

Besides the above filtering on parameter parallax\_over\_error, we have
experimented with the quality filters number (1) to (3) suggested by
Arenou \e (2018). Filter (1) is based on the astrometric $\chi^2$,
modulated by the Gaia $G$ magnitude of the object; filter (2) is a
photometric quality filter, only useful if using Gaia colors; filter (3)
is based on the number of Gaia observations contributing to a given
measurement. It turns out however, that applying filters (1) and (3) to
the Gaia source list obtained above produces a rather large effect, with
only { 149\,141 of the sources ($\sim 48$\%) passing the selection. }
We have therefore tried to estimate whether this refinement brings a worth
improvement in our Sco~OB2 Gaia source sample.
Fig.~\ref{pm-select-init} shows the Gaia source density in the
proper-motion (PM) plane $(\mu_{l},\mu_{b})$ in Galactic coordinates,
with filters (1) and (3) applied. The figure also shows the Gaia PMs of
the probable Sco~OB2 (Hipparcos) members from de~Zeeuw \e (1999), which coincide
with a source density enhancement. The bulk of field sources appear
rather spread across the whole PM plane, with only a modest enhancement
near $(\mu_{l},\mu_{b})=(-5,0)$. The red polygon encloses all Hipparcos
members and the corresponding Gaia-source density enhancement, and
therefore defines our initial member selection region.

By contrast, Fig.~\ref{pm-nofilt} shows the Gaia source density in
the same PM plane, but without filters (1) and (3) from Arenou \e
(2018): the difference is striking, especially in the bulk of field
stars near $(-5,0)$ (outside the initial selection polygon), and at the
same time detailed structures inside the member-selection region are
much better defined. Therefore, filtering with criteria (1) and (3)
appears to be overdone, especially for the field stars, and we choose
not to apply these further restrictions to our initially selected Gaia
sample. It is nevertheless reassuring that the fractional rejection
caused by these additional filters affects likely members (inside the
red polygon) much less than field stars, so that our final results are
qualitatively unaffected by either choice.

Figure~\ref{pm-select-init} shows also the median error on PM, and the
size of the smoothing Gaussian: both are much smaller than the
distinct substructures which are evident in the density distribution of
candidate members: these latter indicate therefore resolved dynamical
structures of the member population, which we will examine in detail in
the next sections.
{
The number of initial candidate members (inside the red polygon) is 40512.
We will refer to them as PM-selected stars.
}

{
\section{Space velocities}
\label{space}

Before proceeding further, it is necessary to recall that a kinematical
stellar group is defined as having a common space motion, rather than a
common proper motion. For a compact group (less than 5-10 degrees on the
sky) there is little difference between the two representations, hence
analysis in the proper motion plane is usually sufficient to study memberships
in compact clusters. This is not true if an association of stars, such
as Sco~OB2 in its entirety, extends over several tens of degrees.
We therefore devote this section to the development of a
{ novel procedure (as far as we known)} able
to characterize diffuse kinematical groups, for each star of which 5D data are
available (position, parallax, proper motion), as in the present case.

The equations to convert sky positions (including distance $R$) and
motion to space coordinates and velocities are:
\begin{equation}
X = R \cos b \cos l
\label{eq1}
\end{equation}
\begin{equation}
Y = R \cos b \sin l
\label{eq2}
\end{equation}
\begin{equation}
Z = R \sin b
\label{eq3}
\end{equation}
then, space velocities $U,V,W$ are:
\begin{equation}
U \equiv \frac{dX}{dt} = \frac{dR}{dt} \cos b \cos l - 
  R \sin b \cos l \; \frac{db}{dt} - R \cos b \sin l \; \frac{dl}{dt}
\end{equation}
\begin{equation}
V \equiv \frac{dY}{dt} = \frac{dR}{dt} \cos b \sin l -
  R \sin b \sin l \; \frac{db}{dt} + R \cos b \cos l \; \frac{dl}{dt}
\end{equation}
\begin{equation}
W \equiv \frac{dZ}{dt} = \frac{dR}{dt} \sin b + R \cos b \; \frac{db}{dt}
\end{equation}
{
These definitions of $U,V,W$ follow the same convention adopted in
many previous works (e.g., de Zeeuw \e 1999, Dehnen and Binney 1998,
Famaey \e 2005, van~Leeuwen 2009)
\footnote{There is no universal convention
on the definition of $U$, with some authors defining it as $U = -dX/dt$,
{ as mentioned by Johnson and Soderblom (1987)}.}.
}
Now, we have (where $v_R$ indicates radial velocity):
\begin{equation}
\frac{dR}{dt} = v_R
\end{equation}
\begin{equation}
\frac{db}{dt} = \mu_b
\end{equation}
\begin{equation}
\cos b \; \frac{dl}{dt} = \mu_l
\end{equation}
so that the above equations transform to:
\begin{equation}
U = v_R \cos b \cos l - 
  R \sin b \cos l \; \mu_b - R \sin l \; \mu_l
\label{eq10}
\end{equation}
\begin{equation}
V = v_R \cos b \sin l -
  R \sin b \sin l \; \mu_b + R \cos l \; \mu_l
\label{eq11}
\end{equation}
\begin{equation}
W = v_R \sin b + R \cos b \; \mu_b.
\label{eq12}
\end{equation}
By simple algebraic manipulations, the latter three equations may be
solved for proper motions and $v_R$, that is:
\begin{equation}
R \; \mu_l = V \cos l - U \sin l
\label{eq13}
\end{equation}
\begin{equation}
R \; \mu_b = W \cos b - (V \sin l + U \cos l) \sin b
\label{eq14}
\end{equation}
\begin{equation}
v_R  = W \sin b + (V \sin l + U \cos l) \cos b.
\label{eq15}
\end{equation}

The term $R \mu_l = V_l$ is nothing else as the transverse velocity along
$l$, and similarly $R \mu_b = V_b$.
In Sco~OB2, $v_R$ is known for only a very small percentage of members.

In the case of a spatially compact population (a few degrees on the sky),
the $\sin l$, $\cos l$, $\sin b$ and $\cos b$ are nearly constant, and
clustering on the PM plane (and parallax) guarantees also clustering in
space velocities. For diffuse populations covering many (tens of) degrees
on the sky, instead, variations in the sine/cosine terms cannot be neglected.
However, the specific (strong) assumption that a group of stars shares the same
space motion ($U=U_0, V=V_0, W=W_0$) can still be tested, even if their
$v_R$
values are not known.
Defining $\xi = \tan l$ and $\digamma = R \mu_l \, /\cos l = V_l \, /\cos l$
($\digamma$ is the greek letter ``digamma''),
both of which are measured for every star, we may rewrite Eq.\ref{eq13} as:
\begin{equation}
\digamma = V_0 - U_0 \, \xi
\label{eq16}
\end{equation}
which is a simple straight line in the $(\xi,\digamma)$ plane, the same for all
stars in a kinematical group.
Conversely, a group of stars not falling along the same straight line
in the $(\xi,\digamma)$ plane cannot be a single kinematical group.
The intercept and slope of the
line provide space-velocity components $V_0$ and $U_0$, respectively. The third
velocity component $W_0$ can then be derived from Eq.\ref{eq14}, suitably
rewritten as:
\begin{equation}
W_0 = \frac{R \; \mu_b + (V_0 \sin l + U_0 \cos l) \sin b}{\cos b}
\label{eq17}
\end{equation}
where $U_0$ and $V_0$ are derived from the previous step, and
all other quantities on the right-hand side are measured for each star.
Unlike $U_0$
and $V_0$, derived from a population best-fit, $W_0$ from Eq.\ref{eq17}
is computed for each star in a given group. Kinematical coherence
requires that the distribution of such $W_0$ values be sharply peaked,
consistently with the expected width of the actual $W$ distribution
(typically 1-2 km/s), and the propagated measurement errors, here dominated
by errors on parallaxes.
We define this method as our method~A.

We have also devised an alternative method to recover $(U_0, V_0, W_0)$ for a
diffuse kinematical population with Gaia measurements, but in the
absence of measured $v_R$.
In fact, in Eqs.~\ref{eq10} to~\ref{eq12}
terms $l,b,R,\mu_l,\mu_b$ are all known, and those equations may be
condensed as:
\begin{equation}
U = v_R \,a_1 + b_1
\label{eq18}
\end{equation}
\begin{equation}
V = v_R \,a_2 + b_2
\label{eq19}
\end{equation}
\begin{equation}
W = v_R \,a_3 + b_3
\label{eq20}
\end{equation}
where $a_i$ and $b_i$ are constants (for each given star). These
parametric equations describe a straight line in the $(U,V,W)$ space, with
$v_R$ as a parameter. For varying $v_R$, each star corresponds to a line.
For a kinematical group of stars, all the corresponding lines will converge
to the common point $(U_0,V_0,W_0)$. This method permits a simultaneous
determination of all three velocity components, unlike method~A above.
We define this second method as method~B.
In its essence, it is analogous to the ``spaghetti method'' from
Hoogerwerf and Aguilar (1999), however in a simplified form, which is
justified by the much higher precision of Gaia measurements compared to
Hipparcos.

\subsection{Expansion motions}

If a kinematical group expands linearly, its space velocities will be:
\begin{equation}
U = U_0 + k (X-X_0)
\label{eq21}
\end{equation}
\begin{equation}
V = V_0 + k (Y-Y_0)
\label{eq22}
\end{equation}
\begin{equation}
W = W_0 + k (Z-Z_0)
\label{eq23}
\end{equation}
where $(X_0,Y_0,W_0)$ is the center of expansion and $k$ is a constant.
We remark that there
is circularity in this definition, since the center of expansion is
defined as the (only) point in space where $(U,V,W) = (U_0,V_0,W_0)$,
but, conversely, $(U_0,V_0,W_0)$ is defined as the space velocity in the
center of expansion. Exactly the same dynamics may be described assuming
$(X,Y,Z) = (0,0,0)$ as the center of expansion, and
$(U_0-k X_0,V_0-k Y_0,W_0-k Z_0)$ as its space velocity.
This is a conceptual ambiguity, and cannot be solved by any analysis
method, as long as the velocity law is linear.
By inserting the velocity laws Eqs.~\ref{eq21}-\ref{eq23} into
Eqs.~\ref{eq10}-\ref{eq12}, combined with definitions in
Eq.~\ref{eq1}-\ref{eq3}, we obtain that
\begin{equation}
U_0-k X_0 = (v_R -kR) \cos b \cos l - 
  R \sin b \cos l \; \mu_b - R \sin l \; \mu_l
\label{eq24}
\end{equation}
\begin{equation}
V_0-k Y_0 = (v_R -kR) \cos b \sin l -
  R \sin b \sin l \; \mu_b + R \cos l \; \mu_l
\label{eq25}
\end{equation}
\begin{equation}
W_0-k Z_0 = (v_R -kR) \sin b + R \cos b \; \mu_b.
\label{eq26}
\end{equation}
With the same manipulations as above, we obtain again linear relations
like Eqs.~\ref{eq13} and~\ref{eq16}, that is
\begin{equation}
R \; \mu_l = (V_0-k Y_0) \cos l - (U_0-k X_0) \sin l
\label{eq27}
\end{equation}
\begin{equation}
\digamma = (V_0-k Y_0) - (U_0-k X_0) \, \xi
\label{eq28}
\end{equation}
which show that even in the case of linear expansion there is a linear
relation between $\xi$ and $\digamma$, in itself indistinguishable from
that found in the absence of expansion. As Eqs.~\ref{eq13} and~\ref{eq27}
show, the observed slope in the diagram $(l,V_l)$ cannot be naively
attributed to expansion or contraction, since any slope can be obtained
for a suitable pair of constants $(U_0-k X_0,V_0-k Y_0)$, with no
constraints on $k$. Therefore,
analysis of the stellar locus in the $(\xi,\digamma)$ plane does not
by itself provide any constraint on the expansion parameter $k$.
This is both bad news and good news: if on one hand the diagram does not
permit to asses if expansion or contraction takes place (i.e., the value
of $k$), on the other hand even a kinematical population with $k\neq 0$
can be recognized as falling along a straight locus in the $(\xi,\digamma)$
diagram.
In the same way, Eq.~\ref{eq17} can be used to compute $W_0-k Z_0$ once
$(U_0-k X_0,V_0-k Y_0)$ are known, but not to derive $k$.
More in general, careful analysis of Eqs.~\ref{eq24}-\ref{eq26} shows that
the determination of $k$ is only possible if $v_R$ is also measured, together
with the other spatial and kinematical parameters. Therefore, only a
complete 6D dataset (not available for the majority of Gaia sources) can
provide us with a measure of $k$. This coefficient is determined with the
help of Eq.~\ref{eq15}, rewritten of course as:
\begin{equation}
v_R -kR = (W_0-kZ_0) \sin b + ((V_0-kY_0) \sin l + (U_0-kX_0) \cos l) \cos b
\label{eq29}
\end{equation}
in whose right-hand side $(V_0-kY_0)$ and $(U_0-kX_0)$ come from the linear
fit to Eq.~\ref{eq28}, and $(W_0-kZ_0)$ from Eq.~\ref{eq17} rewritten as
\begin{equation}
W_0-kZ_0 = \frac{R \; \mu_b + ((V_0-kY_0) \sin l + (U_0-kX_0) \cos l)
\sin b}{\cos b}.
\label{eq30}
\end{equation}
In this way, the quantity $v_R^\prime = v_R -kR$ can estimated for each star,
and individual measurements of $v_R$ and $R$ (from parallax) permit to test
the existence of expansion/contraction from the correlation between
$v_R-v_R^\prime$ and $R$, 
and to compute the coefficient $k = (v_R-v_R^\prime)/R$.
A similar conclusion was reached by Blaauw (1964) and Pecaut \e (2012).

An exception may occur if a compact cluster of stars shows a correlation
between (for instance) $l$ and $V_l$, suggesting expansion or contraction, and
the (absolute) values of $U_0$ or $V_0$ implied by Eq.~\ref{eq27} assuming
$k=0$ would be unreasonably large. In this case a (rough and preliminary)
estimate of $k$ might be obtained by setting $(X_0,Y_0)$ as the cluster
center, and assuming $U_0 \sim V_0 \sim 0$, regardless of $v_R$
measurements.
}

\begin{figure}
\resizebox{\hsize}{!}{
\includegraphics[]{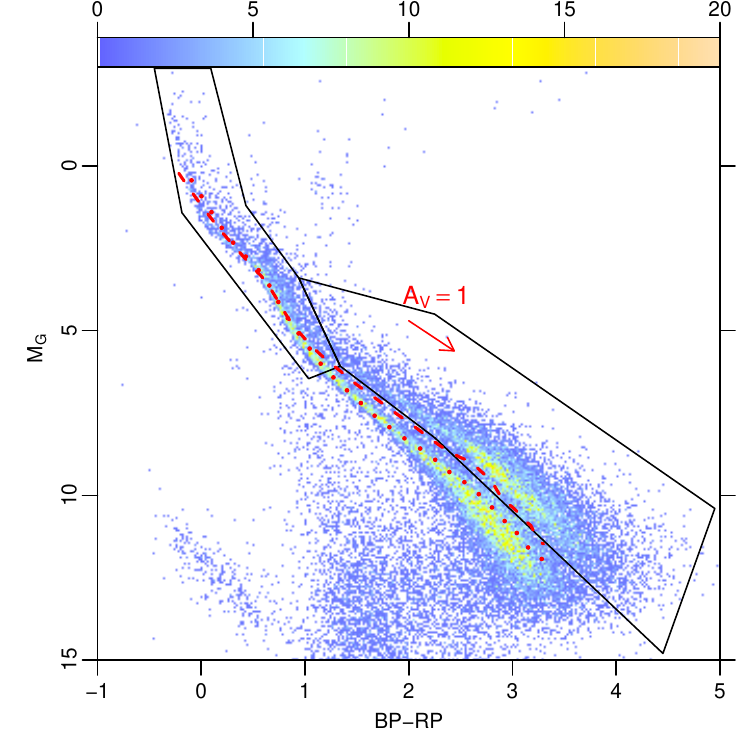}}
\caption{
Color-Absolute Magnitude Diagram (CAMD) of all Gaia sources in our
{ PM-selected} member sample (polygon in Fig.~\ref{pm-nofilt}).
The red arrow is the reddening vector from Kounkel \e (2018).
{
The solid black polygon redward of $BP-RP \sim 1$ defines our ``PMS locus''.
The solid black polygon blueward of $BP-RP \sim 1.4$ defines the ``Upper
MS locus''.
The dotted red line is the Pleiades sequence, while the dashed red
line is the sequence of cluster IC~2602, both from Gaia data.
}
\label{camd-pms-define}}
\end{figure}

\section{Membership to Sco OB2}
\label{members}

Here we examine the various indicators for Sco~OB2 membership provided by the
Gaia data, taking also advantage of the low relative error on parallaxes
in the selected sample.

\subsection{Color-Absolute Magnitude Diagram}
\label{camd}

A useful diagnostic tool enabled by the accurate Gaia data is the
Color-Absolute Magnitude Diagram (CAMD, shown in
Fig.~\ref{camd-pms-define}), based on Gaia BP and RP
photometric measurements, and Gaia absolute $G$ magnitude $M_G$ (see
e.g., Gaia Collaboration \e 2018b).
The error on $M_G$ introduced by parallax errors,
given our selection on $\pi/\Delta \pi$, is 0.22~mag at most, and typically
much less except for the faintest sources.

The CAMD in Fig.~\ref{camd-pms-define} shows only Gaia sources in the
{ PM-selected} member sample (red polygon in Fig.~\ref{pm-nofilt}).
This diagram shows both a well-defined main sequence (MS), and a clear
pre-main-sequence (PMS) band above it, separated by a gap.
Also shown for reference are the empirical sequences of the Pleiades
and IC~2602\footnote{The IC~2602 sequence is computed from all Gaia
members selected from a square sky region of $5^\circ$ side centered on
$(RA,Dec)=(160.74167,-64.4)$, and with
constraints $\pi/\Delta \pi>10$, $-26<\mu_l<-17$~mas/yr, $-2<\mu_b<4$~mas/yr,
and $6<\pi<7$~mas (383 stars).} clusters, derived by us using Gaia DR2 data.
{ In the lower left part ($BP-RP<1$, $M_G>10$) the white-dwarf sequence
is also visible.}
The low-density,
diffuse cloud of datapoints below the MS at colors $BP-RP>1$ is due to
the so-called ``$BP-RP$ excess'' (e.g., Arenou \e 2018), arising from
source confusion in dense areas, for which the $BP$ and $RP$ colors are
unreliable. In our sample this effect involves a
negligible fraction of Gaia sources, even though we did non apply
filter (2) from Arenou \e (2018), purposely designed to remove those
sources. The PMS band comprises all the low-mass members of Sco~OB2,
while the more massive MS members ($BP-RP<1$) cannot be distinguished from
the field stars from this CAMD. The red arrow is the reddening vector,
as estimated by Kounkel \e (2018) for low extinction values: the narrow
width of the MS at $BP-RP<1$ (i.e., where it is not parallel to the
reddening vector) shows that extinction is negligible up to the maximum
sample distance of 200~pc. Exceptions to this are only likely in the
immediate vicinity of dark clouds ($\rho$~Oph, Lupus), which occupy
a tiny fraction of the studied sky area.
The { larger} solid black polygon in Fig.~\ref{camd-pms-define} encloses all
possible PMS stars up to very young ages, and at the same time is
affected by only a minimal contamination from low-mass MS field stars,
thanks to the MS narrow width: this polygon defines our ``PMS locus''
in the CAMD, and contains { 10839} sources (strongly dominated by members),
reported in Table~\ref{table-pms} { and henceforth referred to as PMS
sample}.
The oldest ages covered by this PMS locus correspond approximately to
the age of IC~2602 ($\sim 45$~Myr, e.g., Dobbie \e 2010).
The upper-MS locus of possible Sco~OB2 members falls instead
inside the { smaller} black { polygon}, in the upper left part of
the diagram, and contains { 3598} sources (both members and field stars),
reported in Table~\ref{table-upp-ms} { and henceforth referred to as
Upper-MS sample.
{
In both Tables~\ref{table-pms} and~\ref{table-upp-ms} we also
include star identifiers from SIMBAD, when available;
the match between the SIMBAD database and the Gaia DR2 catalog was made
by CDS (Strasbourg) and not checked by us. Out of the 10839 PMS members
in Table~\ref{table-pms} only 1840 (17\%) have a corresponding entry in
SIMBAD; instead, a SIMBAD match is found for 2922 (81\%) of the 3598
Upper-MS members in Table~\ref{table-upp-ms}.
}
The PMS and Upper-MS samples will be referred to collectively as the CAMD
candidate member sample (14437 stars).
The CAMD of Fig.~\ref{camd-pms-define} also shows a modest number of
red-clump giants, and an even smaller number of brighter and redder giants
at $BP-RP>2$ and $M_G<0$. This type of stars cannot be a contaminant for
the CAMD candidate sample.
}

\begin{figure}
\resizebox{\hsize}{!}{
\includegraphics[]{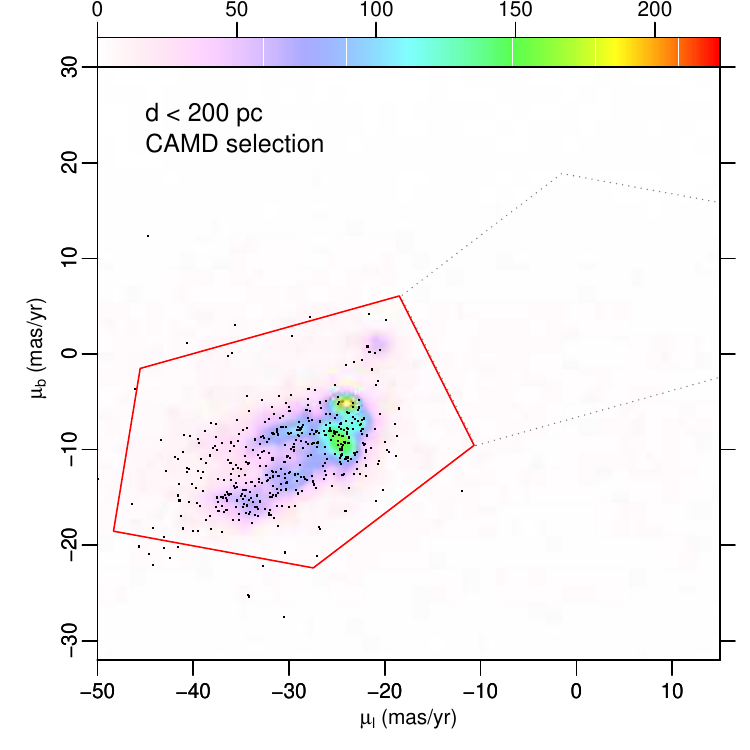}}
\caption{
{
Same as Fig.~\ref{pm-nofilt} (with the same color scale), for all stars
in the PMS and Upper-MS loci from Fig.~\ref{camd-pms-define}.
The red polygon is the same as in Fig.~\ref{pm-nofilt}.
Dots indicate members from PM16 (their Table~7), with Gaia
parallax $\pi>5$.
Outside the red polygon the density of stars is very low, but not zero.
The dotted grey polygon is a reference region to estimate contamination
(see Sect.~\ref{contamin}).
}
\label{pm-pms}}
\end{figure}

\subsection{Proper-motion Diagram}
\label{pm}

The combination of photometric selection from the CAMD with astrometric
selection from the PM plane provides us with a very clean sample of
Sco~OB2 members, especially for the (numerically dominant) low-mass PMS
stars,
{
less cleanly for the Upper-MS stars. In fact,}
Fig.~\ref{pm-pms} shows the density of { CAMD candidates (with no
constraints on PM)}
in the PM plane, as in Fig.~\ref{pm-nofilt}.
{
The CAMD selection has the obvious effect of removing the largest
majority of field stars, both near $(\mu_l,\mu_b) \sim (0,0)$, and
diffusely across the PM diagram (Figs.~\ref{pm-pms} and~\ref{pm-nofilt}
have the same color scale).
We estimate the residual contamination from field stars in the
CAMD$+$PM-selected sample (i.e., the stars shown in Fig.~\ref{pm-pms}
and falling inside the red box) in Sect.~\ref{contamin} below.
}
The density distribution of { CAMD candidates}
in the { PM} plane is highly structured. The comparison with median
errors and width of the smoothing Gaussian kernel shows that all those
structures are real, up to a level of detail even greater than that
shown in Fig.~\ref{pm-nofilt}.
{
Fig.~\ref{pm-pms} also shows the Gaia DR2 proper-motion data (small dots)
of stars classified as Sco~OB2 members by PM16 (their
Table~7, henceforth ``PM16 members''), matched with Gaia positions
within 1~arcsec.  Of the 493 PM16 members, 454 have a successful Gaia
DR2 match, 451 with $\pi>5$. Thus, only 3/454 of matched PM16 members lie
farther out than 200~pc, a much smaller percentage than the de~Zeeuw \e
(1999) massive candidate members ($\sim 8$\%, see Sect.~\ref{obs}); this
reinforces our arguments for a limiting distance of 200~pc in the
present study.
The vast majority of the PM16 members in Fig.~\ref{pm-pms} follows the same
density pattern as our CAMD candidate members, as expected.
However, of the 451 PM16 members with $\pi>5$, 18 (4\%) fall outside our
PM-selection region (the red box in Fig.~\ref{pm-pms}). There may be a
variety of reasons for this, one of which may be their nature as
astrometric binaries, which are not classified as such in Gaia DR2. If
so, and assuming that all PM16 members are true members, we should conclude
that our PM selection misses of order of 4\% of true members. However,
enlarging the box further to recover them is at risk of including too
many contaminants (see below) and this option will not be considered
further.

\begin{figure}
\resizebox{\hsize}{!}{
\includegraphics[]{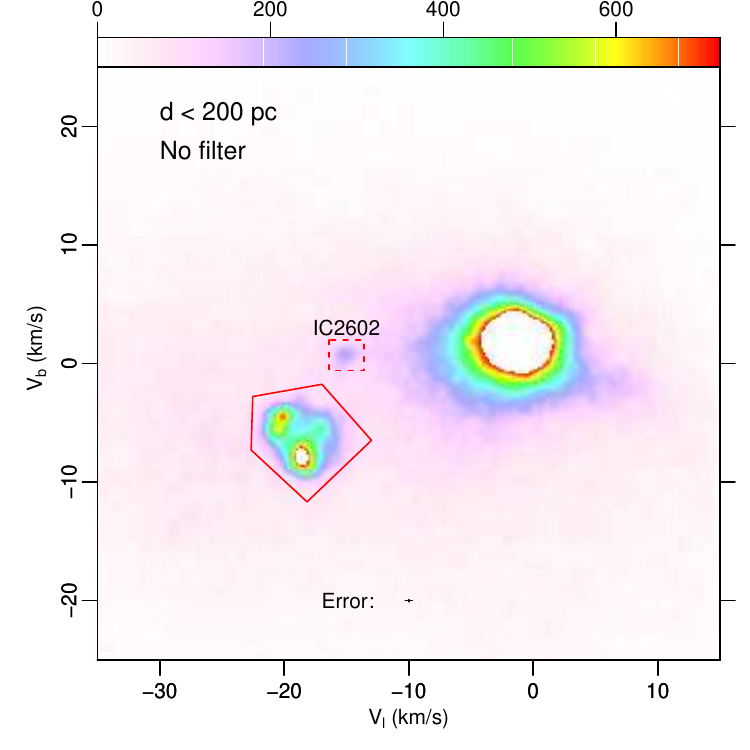}}
\caption{
{
Transverse-velocity plot $(V_l, V_b)$, for the same sample as in
Fig.~\ref{pm-nofilt}. The median error is indicated by black segments.
The solid red box encloses the majority of Sco~OB2 members and was chosen
independently from that in Fig.~\ref{pm-nofilt}.
The dashed red box encloses stars in IC~2602.
}
\label{vt-all}}
\end{figure}

\begin{figure}
\resizebox{\hsize}{!}{
\includegraphics[]{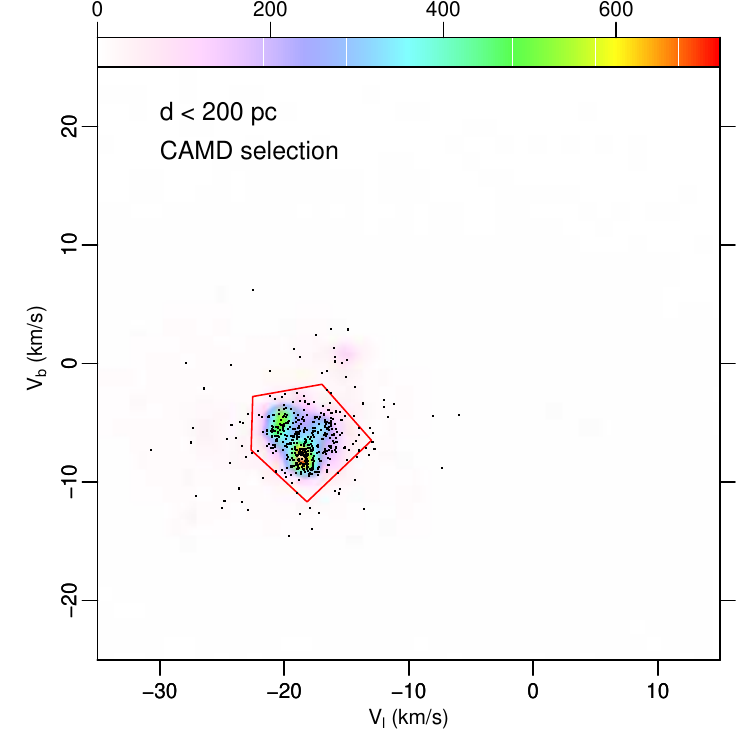}}
\caption{
{
Transverse-velocity plot $(V_l, V_b)$, for PMS and Upper-MS stars as in
Fig.~\ref{pm-pms}. The red box is as in Fig.~\ref{vt-all}.
Black dots are members from PM16 as in Fig.~\ref{pm-pms}.
}
\label{vt-pms}}
\end{figure}

\subsection{Transverse velocities}
\label{vt}

Some of the structures in the CAMD member distribution in Fig.~\ref{pm-pms} are
in all probability related to the wide distribution of these stars
in the sky, which involves non-negligible projection effects
(Sect.~\ref{space}). No less important than the apparent (sky-projected)
space distribution is the depth distribution, suggested by a correlation
(not shown for brevity) which we find between proper motion and parallax.
In cases where the true space-velocity vector is close to normal to the line of
sight, depth may become the dominant factor in the spread across the PM
plane for association members. 
To test this, we have computed transverse velocities (VT)
$V_l$ and $V_b$ (Sect.~\ref{space}), whose distribution is shown in
Fig.~\ref{vt-all} for our entire sample, and in Fig.~\ref{vt-pms} for
the CAMD-selected sample.
It is obvious that the candidate Sco~OB2 members distribution is here
more compact compared to that in the PM plane of Fig.~\ref{pm-pms},
indicating that depth effects do indeed play a major role on the
apparent motion of association members.
The error on $V_l$ and $V_b$ is often dominated by errors on $\pi$
rather than on proper motion; we recall however that our sample was required to
satisfy $\pi/\Delta \pi >10$, so that, on average, errors on transverse
velocities are small, as shown in the Figure, and velocity structures in
the plot are real.
Therefore, depth effects are not uniquely responsible for the apparent
dynamical structures, and it is legitimate to investigate about other
projection effects.
The red polygon shown in Fig.~\ref{vt-all} and Fig.~\ref{vt-pms} encloses the
majority of Sco~OB2 members and was chosen independently from that in
Fig.~\ref{pm-pms}. It represents a more conservative selection
compared to Fig.~\ref{pm-pms}. This becomes clear from a quantitative
comparison between Fig.~\ref{vt-pms} and Fig.~\ref{pm-pms}: in the
former the red region encloses 11058 CAMD-selected
members (henceforth VT-selected members), compared to 14437 candidate
CAMD members in the latter.

Fig.~\ref{vt-pms} also shows the PM16 members of Sco~OB2: of 451 stars
with $\pi>5$, 69 (15.3\%) fall outside our selection (red polygon). We
conclude that the VT-selected sample is a highly reliable, but
significantly incomplete selection of association members; in the
following, we will consider this sample uniquely to address questions
where the highest astrometric quality is involved. For general
membership issues, the VT-selected sample is too much incomplete, and we
prefer to consider the PM selection from Fig.~\ref{pm-pms}.

Table~\ref{table-breakdown} summarizes the sample statistics, by sub-region
and adopted criteria. Numbers in column ``All'' are not the sum of the
three preceding columns, since a (small) number of candidate members
lie outside the adopted boundaries of USC, UCL and LCC regions.
From a comparison between Figs.~\ref{pm-pms} and~\ref{vt-pms} it is also
clear than the VT selection leaves out IC~2602 stars, unlike PM
selection. The number of IC~2602 stars from the VT diagram (and CAMD
selection) is of 253 PMS and 114 Upper-MS stars, which should be added
to column ``LCC'' of Table~\ref{table-breakdown}, rows 3-4, for a more
accurate comparison with rows 1-2. The Table shows that VT selection
lowers the number of PMS candidates by 10-20\%, while it operates a much
more drastic reduction ($>50$\%) on the number of Upper-MS candidates,
which is again indication of the higher non-member contamination in the
Upper-MS subsample.

\begin{figure*}
\resizebox{\hsize}{!}{
\includegraphics[]{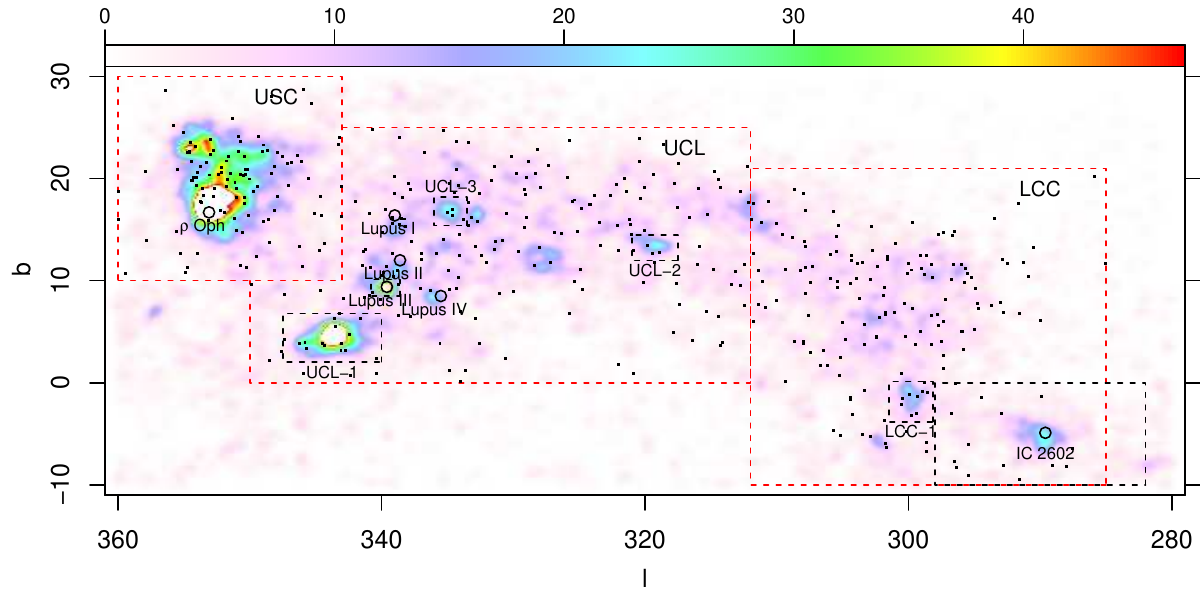}}
\caption{
Spatial density of { PM-selected} PMS stars.
The top-axis color bar indicates stellar density in units of stars per square
degree.
Pixels with densities above the color-scale maximum are shown in white.
The large red dashed rectangles indicate the boundaries of the Upper
Sco-Cen (USC), Upper Centaurus-Lupus (UCL), and Lower Centaurus-Crux
(LCC) regions, after de~Zeeuw \e (1999) and Preibisch and Mamajek (2008).
Black dots are members after de~Zeeuw \e (1999).
Positions of known star-forming regions from Mellinger (2008), and of
the open cluster IC~2602, are indicated with labeled small circles.
The dense cluster around V1062~Sco is labeled UCL-1.
{
This and the other small clusters UCL-2, UCL-3, LCC-1, as well as
IC~2602 and the Lupus~III condensation are indicated by dashed
rectangles.
}
\label{spatial-map-all-satur}}
\end{figure*}

\begin{table}
\centering
\caption{Summary of candidate Sco~OB2 members}
\label{table-breakdown}
\begin{tabular}{lrrrr}
  \hline
                & USC  & UCL  & LCC  & All  \\
   \hline
PM $+$ PMS      & 2862 & 4511 & 2803 & 10839 \\
PM $+$ Upper-MS &  441 & 1261 & 1208 &  3598 \\
VT $+$ PMS      & 2587 & 4077 & 2154 &  9221 \\
VT $+$ Upper-MS &  215 &  574 &  381 &  1337 \\
   \hline
\end{tabular}
\end{table}

}

\begin{figure*}
\resizebox{\hsize}{!}{
\includegraphics[]{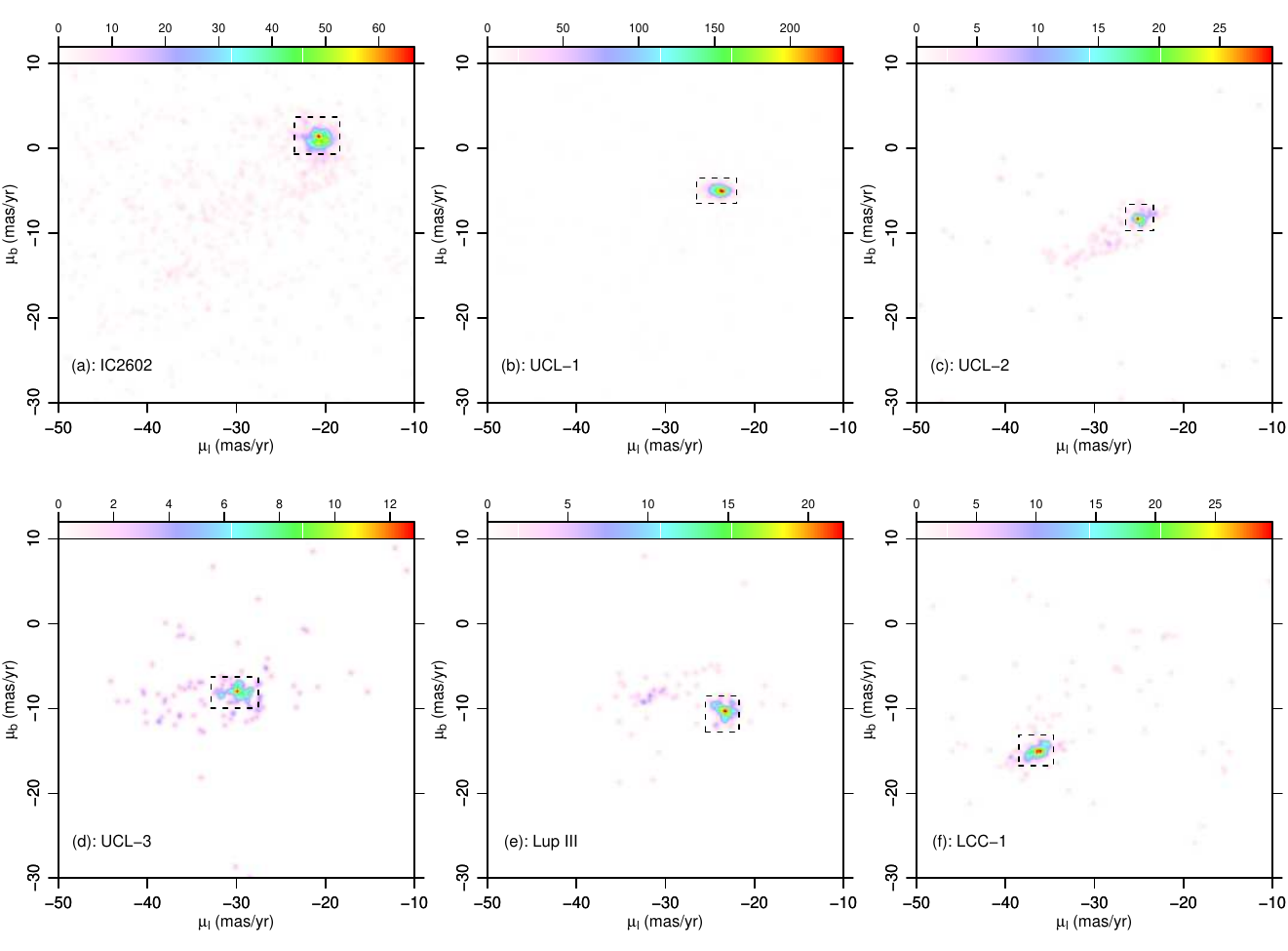}}
\caption{
{
PM diagrams of spatially compact populations (PMS and Upper-MS), as
defined by dashed rectangles in Fig.~\ref{spatial-map-all-satur}, and
no other constraints on PM.  Only stars inside the dashed rectangles in
the PM plane are kept as members of the respective compact populations.
}
\label{pm-compact}}
\end{figure*}

{
\subsection{Contamination}
\label{contamin}

From the previous Section it is qualitatively clear that field-star
contamination in the PM$+$CAMD sample is very small (and even less in
the VT$+$CAMD sample). Here we try to quantitatively estimate this
contamination. Since the spatial region we study is very large, it makes
little sense to look for a comparably large sky region, where one might
reasonably assume
to find the same distribution of Galactic field stars but total absence
of PMS stars. A reference set of measurements where to evaluate
contamination can be instead found in the PM diagram, still considering
the same spatial region of our whole sample. In particular, we have
shown that the largest majority of field stars across Sco~OB2 (and for
$\pi>5$) are clustered around $(\mu_l, \mu_b) \sim (0,0)$ in
Fig.~\ref{pm-nofilt}, and that CAMD selection rejects virtually all of
them (Fig.~\ref{pm-pms}). If we displace our selection region on the PM
plane (the red polygon in Fig.~\ref{pm-pms}), and re-center it to
$(0,0)$ (with a 180-degree rotation to avoid overlap with the current
selection), we find 344 field
stars inside the grey polygon in Fig.~\ref{pm-pms},
also falling in the PMS region in
the CAMD. This is the absolute maximum contamination which may potentially
affect our PM$+$CAMD member sample: therefore, a maximum contamination
of 3.2\%. For the Upper-MS sample, the same procedure gives both a much larger
number and percentage of contaminants: 1162 field stars, or 32.3\%,
which was qualitatively expected.

We have also estimated contamination using less extreme hypotheses. For
example, by rotating the red polygon in Fig.~\ref{pm-pms} by 90, 180 and
270~deg.\ around $(0,0)$, we obtain respectively 200, 62, and 72
contaminants for the PMS sample (0.6\% to 1.8\%), and 1068, 345, and
275 contaminants for the Upper-MS sample (7.6\% to 29.7\%). Therefore,
a rather robust conclusion is a contamination level between 1-3\% for
the PMS sample, and between 10-30\% for the Upper-MS sample.

We have also considered contamination of the VT-selected samples from
Fig.~\ref{vt-pms}, using the same approach. By translating the red
polygon to $(0,0)$ (absolute-maximum case) we would select 94 field
stars in the PMS sample, and 205 in the Upper-MS sample (1\% and 15\%
contamination, respectively, of the actual VT-selected PMS and Upper-MS
samples). By rotating the selection region around $(0,0)$ we have
36, 4, and 16 PMS contaminants (0.04\% to 0.4\%) and 249, 41, and 47
Upper-MS contaminants (3\% to 18.6\%). As expected, VT selection
involves a significantly lower contamination than PM selection, at the
expense of a significantly lower completeness as seen above.

Therefore, the level of contamination in these Gaia member samples are
much lower than in earlier astrometric studies such as Hoogerwerf
(2000), where contamination ranges from $\sim 30$\% up to near $\sim
60$\%, even considering our worst case of PM-selected Upper-MS members.
}

\section{Spatial distribution of members}
\label{spatial}

The sky-projected spatial density of { PM-selected PMS stars is shown
in Fig.~\ref{spatial-map-all-satur}. }
The total sky area of the three regions is
1974 sq.deg., and the corresponding space volume up to the maximum
distance surveyed here (200~pc) is 1\,653\,883 cubic pc.
The total number of Gaia sources (after our selection on $\pi$ and
$\pi/\Delta \pi$ only) in the three regions is { 164042}, clearly dominated
by field stars: we will discuss in Sect.~\ref{concl} the local space
density ratio between members and field stars.

The highest density peaks of PMS stars in
Fig.~\ref{spatial-map-all-satur}
($\sim 90$ stars/square degree) are found near
the $\rho$~Oph dark cloud, and rather surprising also in correspondence
to a group of stars near V1062~Sco, recently discovered by R\"oser \e (2018)
using Gaia DR1 data, which is labeled as UCL-1 in the Figure.
The global distribution of the new PMS members follows rather closely that of
more massive members from de~Zeeuw \e (1999), also shown in the Figure.
Among the Lupus clouds, we find a noticeable density peak only in
Lupus~III,
{
and a much weaker one in Lupus~IV.
}
Besides the main density peak in $\rho$~Oph, a complex spatial
structure is found all over the Upper~Sco region. At the opposite
extreme, we find a weaker and smoother density peak of PMS stars
corresponding to the ZAMS cluster IC~2602: this is not surprising, since
at the cluster age ($\sim 45$~Myr, e.g., Dobbie \e 2010) the lowest-mass
stars are still found in the PMS stage, and were therefore selected using our
CAMD.
Had we included all IC~2602 MS members, its spatial density peak
would have been much higher.

Figure~\ref{spatial-map-all-satur} provides many details on the
lower-density populations in Sco~OB2. In the densest USC region, density
exceeds 15~stars/sq.deg. over a contiguous region of approximate size
$15^\circ \times 15^\circ$, and exceeds 25~stars/sq.deg. over more than
one-third of the same area. There are no recognizable high-density
condensations bridging the gap between this dense USC region and density
peaks in UCL, like those in the Lupus clouds and the UCL-1 cluster
mentioned above. Apart from Lupus~III, other subclusters in Lupus (Lupus
I, II
and~IV) are rather weak from the Gaia data, compared to e.g., the map in
Preibisch and Mamajek (2008; their Fig.6), and definitely weaker than other
subclusters in UCL, labeled as UCL-2 and UCL-3 in
Fig.~\ref{spatial-map-all-satur}.
The mismatch between the older member map in Preibisch and Mamajek
(2008) and the one in Fig.~\ref{spatial-map-all-satur} should not be
surprising, since the former resulted from observations which were
either spatially incomplete, or with highly non-uniform depth, like the
X-ray observations in Lupus, while the Gaia data are uniform over the
entire region. On the other hand, Gaia is not very sensitive to highly
extincted objects, which are instead more efficiently detected using
X-ray observations.

Another density peak is found in LCC, labeled as LCC-1 in the figure
(containing 90 stars), together with weaker, unlabeled peaks.
Besides these density peaks, the entire Sco~OB2 association is permeated
by a diffuse population of low-mass PMS members with densities of
5-10~stars/sq.deg., running without apparent discontinuity through the
entire length of the association ($\sim 70^\circ$ on the sky).

{
The determination of kinematical groups may be
severely affected by projection effects when a diffuse population spans
several tens of degrees on the sky, as here. This does not hold for
compact clusters, however. Therefore, we separate the study of compact
and diffuse populations, and of their dynamics.

\begin{figure*}
\resizebox{\hsize}{!}{
\includegraphics[]{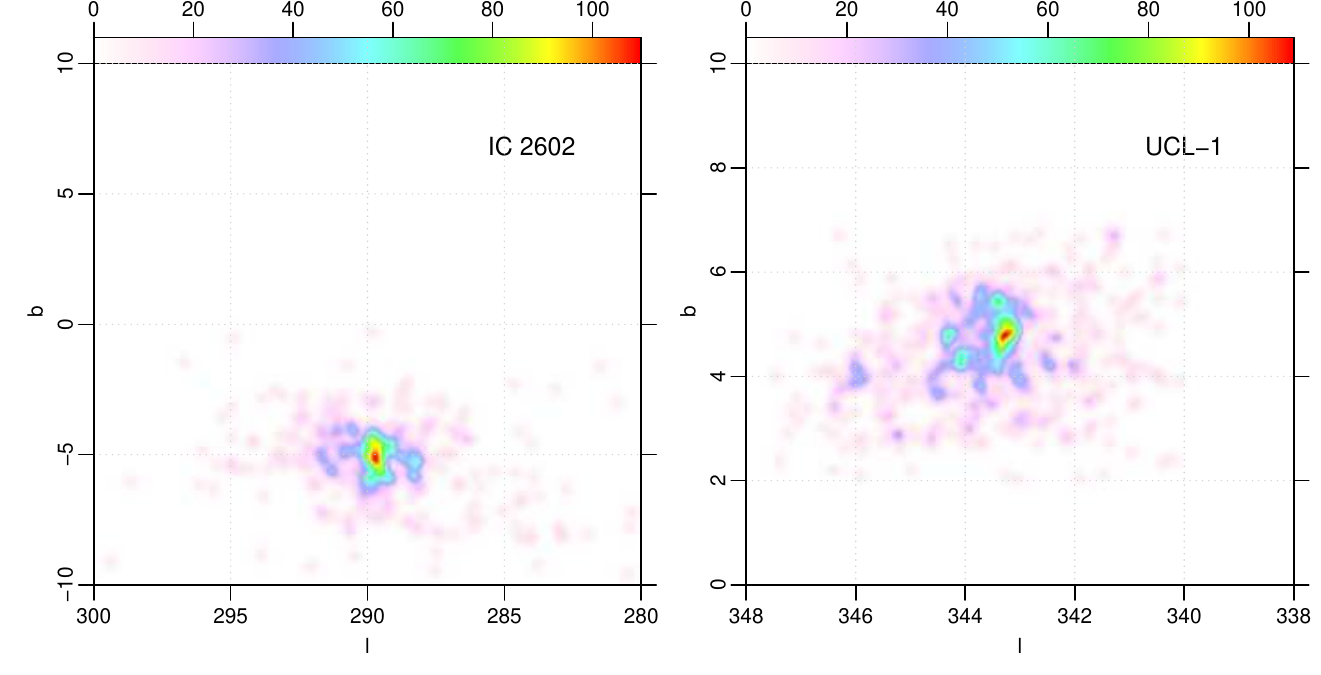}}
\caption{
{
Spatial distributions of IC~2602 (left) and UCL-1 (right) members.
The spatial scale of the two panels is different.
}
\label{ic2602-ucl1-space}}
\end{figure*}

\begin{figure}
\resizebox{\hsize}{!}{
\includegraphics[angle=90]{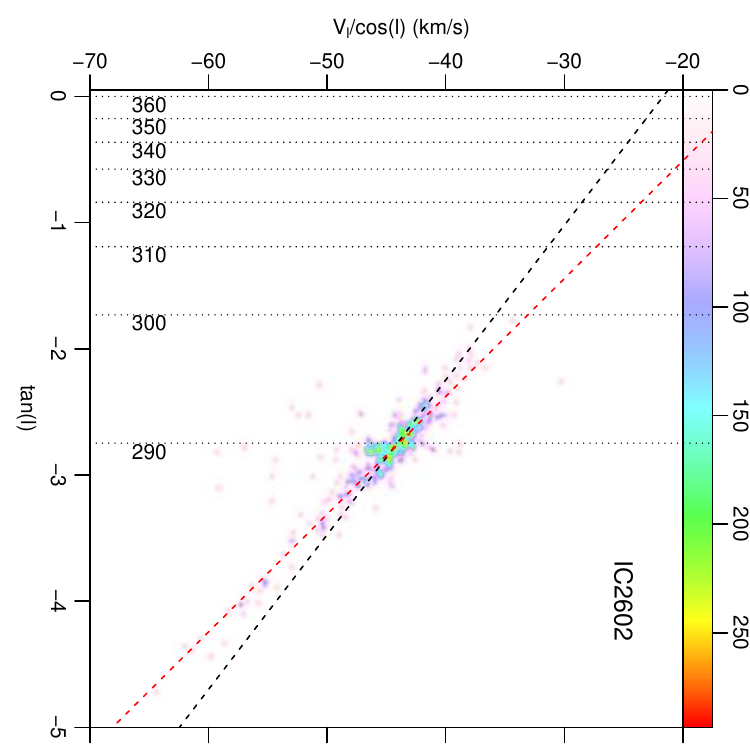}}
\caption{
{
$(\digamma, \xi)$ plot for IC~2602.
The dashed black and red lines are predicted loci of IC~2602 members,
respectively for no expansion, and for expansion with $k=0.05$ km/s/pc.
}
\label{ic2602-xidig}}
\end{figure}

\subsection{Compact populations}
\label{compact}

In the simplest case of a compact (on the sky) population, $b \sim
const$ and $l \sim const$, and Eqs.~\ref{eq13} to~\ref{eq16} show that
$V_l \sim const$ and $V_b \sim const$ (provided of course that the
cluster is not elongated along the line of sight, i.e., $R \sim const$,
as it seems reasonable). In this case projection effects are not
important. Condition for the existence of a clustered and kinematically
coherent population is therefore clustering both in space and on the PM
or VT plane. We have selected from the spatial distribution of members
shown in Fig.~\ref{spatial-map-all-satur} the local overdensities which
most likely correspond to physical groups, well above local density
fluctuations. They are indicated with dashed rectangles in the Figure.
The definition of the USC compact population will be discussed below.

Figure~\ref{pm-compact} shows the distribution on the PM plane of
compact (CAMD-selected) groups, as defined by spatial regions in
Fig.~\ref{spatial-map-all-satur}.  These distributions confirm that these
spatial groups do correspond to kinematically coherent populations,
with most stars in each group being well localized in PM space. We
consider as confirmed members in each group only stars falling inside
the dashed boxes in Fig.~\ref{pm-compact}. Stars from a given spatial
group which fall outside of the corresponding PM box are rejected as
members of the group. We find in this way 350 members (PMS$+$Upper-MS)
in IC~2602, 593 members in UCL-1, 52 in UCL-2, 52 in UCL-3, 90 in LCC-1,
and 69 in Lupus~III.

The spatial distribution of the UCL-1 population is shown in
Fig.~\ref{ic2602-ucl1-space}. 
As already evident from Fig.~\ref{spatial-map-all-satur}, this cluster
is not symmetric, being not only elongated along $l$, but also
possessing a small ``satellite'' cluster (2 degrees
to the East). This small companion cluster was not detected in the discovery
paper (R\"oser \e 2018). A slight asymmetry was also noticeable in the PM
plane (Fig.~\ref{pm-compact}). UCL-1 has a likely complex internal
dynamics, which deserves further studies.

\subsubsection{IC~2602}
\label{ic2602-compact}

Figure~\ref{ic2602-ucl1-space} shows also the spatial distribution of
PMS and Upper-MS members of IC~2602 (not the complete cluster
population, which should also include lower-MS members).
The figure shows the presence of an extended halo
($\sim 10-15^{\circ}$) around this cluster.
The halo density degrades smoothly away from the cluster core, ruling
out a substantial contamination by field stars.
The halo has a radius much larger than the known cluster size
($1.5^\circ$ in radius after
Kharchenko \e 2013), and is asymmetric, being elongated along $l$.
Although IC~2602 does not belong to Sco~OB2, it is
spatially and dynamically a close relative, and for this reason we
include it in this work, if even marginally. The IC~2602 halo
might be populated by cluster members which
are gradually lost (evaporated), as is expected for all but the most
tightly bound clusters (Lada and Lada 2003).
While evaporating, members keep enough memory of their original
kinematics that they still contribute to the same peak in PM space as
stars in the IC~2602 core. 
The Gaia data are therefore providing us with one rather clear detection
of evaporation from a ZAMS cluster, which deserves a deeper study in a
future work.

We have examined if the Gaia data contain indications of a measurable
expansion rate for IC~2602. We have clarified in Sect.~\ref{space} that
an unambiguous determination of expansion requires knowledge of radial
velocities $v_R$. We lack $v_R$ for the new Gaia members, but since
IC~2602 is a well studied cluster we know $v_R$s for members in its
core, as recently measured by Bravi \e (2018) in the context of the
Gaia-ESO Survey: from that work, the median $v_R$ is 17.63~km/s.
The Gaia data provide instead median transverse velocities of $V_l=-15.02$ and 
$V_b=0.77$~km/s. Inserting these values into Eqs.~\ref{eq10} to~\ref{eq12}
we compute median space velocities $(U,V,W)$, which enable us to predict
the locus occupied by the IC~2602 members in the $(\digamma, \xi)$
diagram (Eq.~\ref{eq16}), in the absence of expansion.
The resulting prediction is compared to the data in
Fig.~\ref{ic2602-xidig}. If we instead use Eqs.~\ref{eq24} to~\ref{eq26}
and~\ref{eq28}, and assume an expansion rate of $k=0.05$ km/s/pc, we
obtain a distinctly different locus on the $(\digamma, \xi)$ plane (red
line in the figure). The actual Gaia data lie somewhat in between these
two cases, and deciding between them would involve a detailed membership
assessment (especially) for the halo members, which requires additional
data. The value $k=0.05$ km/s/pc may nevertheless be considered as
a robust upper limit for the IC~2602 expansion rate.
This is consistent with the known age of the cluster: a star in the
halo, $6^{\circ}$ (16~pc) away from the cluster core and traveling at
constant speed since 45~Myr must have a transverse speed of 0.35~km/s, while
its predicted speed with $k\leq 0.05$ km/s/pc would be $\leq 0.8$~km/s,
consistent with this value.

\begin{figure}
\resizebox{\hsize}{!}{
\includegraphics[]{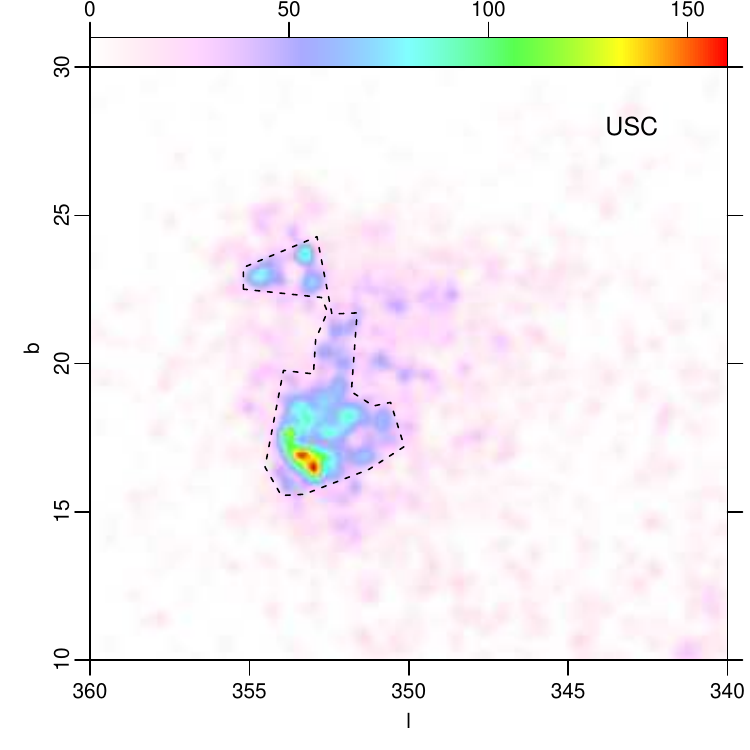}}
\caption{
{
Spatial map of { PM-selected} PMS stars in the USC region.
The dashed line encloses the bulk of the compact component in this area.
}
\label{usc-compact-def}}
\end{figure}

\begin{figure}
\resizebox{\hsize}{!}{
\includegraphics[]{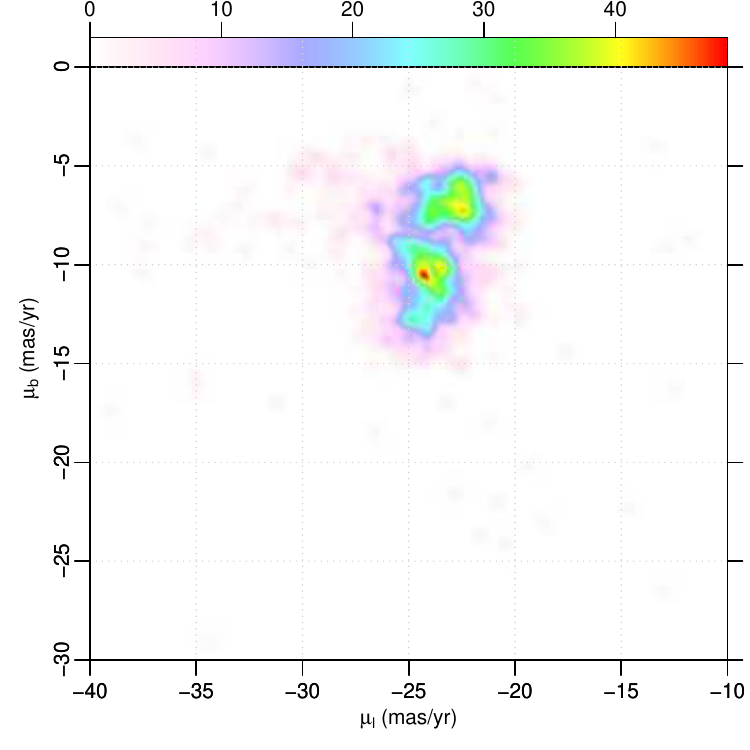}}
\caption{
{
PM diagram of the compact USC component, as defined in
Fig.~\ref{usc-compact-def}.
}
\label{usc-compact-pm}}
\end{figure}

\begin{figure}
\resizebox{\hsize}{!}{
\includegraphics[]{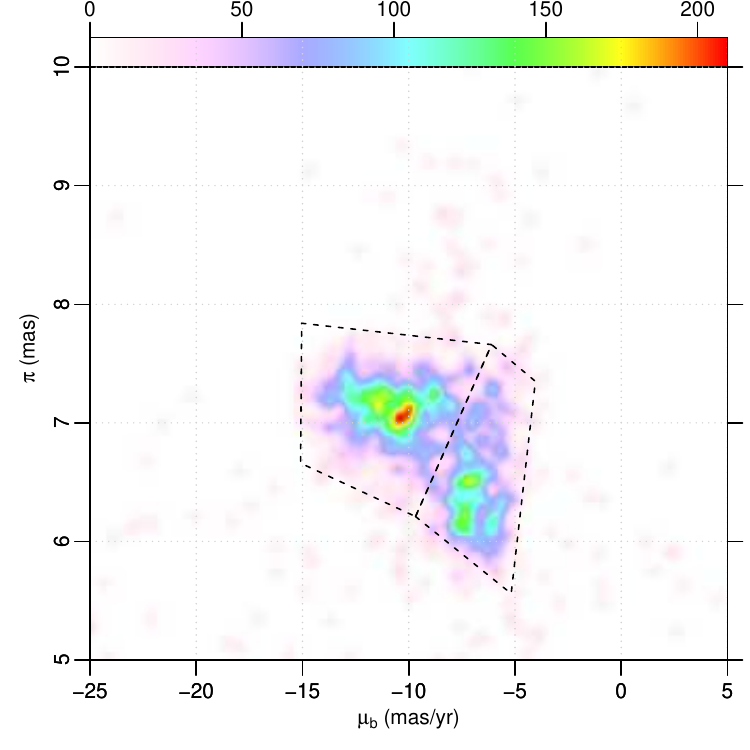}}
\caption{
{
$(\mu_b, \pi)$ diagram of the compact USC component.
The dashed lines select the ``USC near'' and ``USC far'' subpopulations.
}
\label{usc-compact-pm-plx}}
\end{figure}

\subsubsection{Clustered populations in Upper Sco}
\label{usc-compact}

\begin{figure*}
\sidecaption
\includegraphics[width=12cm]{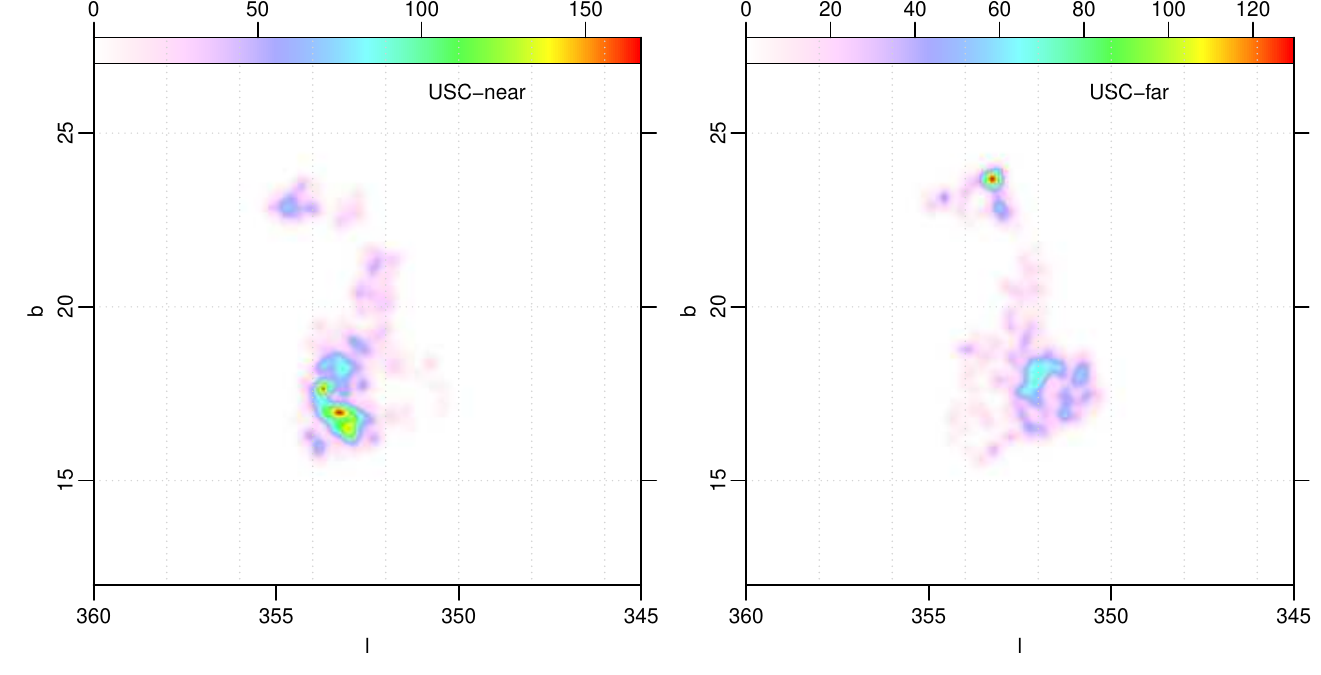} 
\caption{
{
Spatial distributions of ``USC near'' and ``USC far'' subpopulations.
}
\label{usc-compact-space}}
\end{figure*}

In the USC region the distinction between the diffuse and clustered
populations is much less clear-cut than elsewhere in Sco~OB2. As
Fig.~\ref{spatial-map-all-satur} shows, the strongest density peaks in
USC are surrounded by intermediate-density regions ($\sim 30$ stars/sq.degree)
and not immediately by low-density regions ($\leq 10$ stars/sq.degree)
as found throughout most of Sco~OB2. Moreover, the high-density regions in
USC do not possess regular shapes, which makes their definition even
more problematic.
Therefore, we devote this subsection to the definition of the clustered USC
population, and to the study of its peculiarities.

Figure~\ref{usc-compact-def} is a spatial map of all PM-selected PMS
members in the USC region (we avoid here Upper-MS candidates because of
their larger field-star contamination). The map shows an interconnected
aggregate of density peaks, of which the highest roughly corresponding
to the $\rho$~Oph dark clouds. It might be naively suspected that
spatially adjacent peaks share the same dynamics (proper motion), but we
found this not to be true. The spatial region is small enough (less than
$10 \times 10$ sq.degrees as far as the high-density regions are
concerned) that projection effects cannot be responsible for the
difference. We have therefore selected all possible high-density peaks
as clustered-population candidates, and have then examined their dynamics.
The selection was made visually, and corresponds to the dashed line in
Fig.~\ref{usc-compact-def}. It defines 1045 PMS stars, without
constraints on PM or VT. This compact USC population is distributed on
the PM plane as shown in Fig.~\ref{usc-compact-pm}: two main groups are
evident, plus secondary peaks within each of them. The correspondence
between position in space and in the PM plane is not trivial: for
example, the three small, well-defined peaks around $(l,b) \sim
(353,+23)$ do not contribute to the same group in PM space, despite
their proximity on the sky. The same is found for stars near $\rho$~Oph.
The two groups in Fig.~\ref{usc-compact-pm} are better separated along
$\mu_b$ than along $\mu_l$; therefore, we consider also their
distribution on the $\pi, \mu_b$ diagram, shown in
Fig.~\ref{usc-compact-pm-plx}. This shows that, on average, the two
groups have significantly different parallaxes, confirming that the two
groups seen in the PM plane of Fig.~\ref{usc-compact-pm} are real.
Therefore, we conclude that there are two
rather distinct compact populations in USC, which we will refer to as
``USC-near'' ($\pi \sim 7$) and ``USC-far'' ($\pi \sim 6-6.5$).
Operationally, we define their members by using the two dashed regions in
Fig.~\ref{usc-compact-pm-plx}, in addition to the spatial region of
Fig.~\ref{usc-compact-def}. It should be emphasized that neither of the two
subpopulations has a regular structure, there being significant
substructures in both PM space and parallax.
Figure~\ref{usc-compact-space}, finally, shows that substructures in
USC-near and USC-far are also obviously present in their sky
distribution. It is intriguing that within USC close proximity on the sky
does not generally mean belonging to the same kinematical population.
The number of PMS (Upper-MS) members in USC-near and -far are
respectively 501 (19) and 350 (12).

\subsection{Diffuse populations}
\label{diffuse}

\begin{figure}
\resizebox{\hsize}{!}{
\includegraphics[]{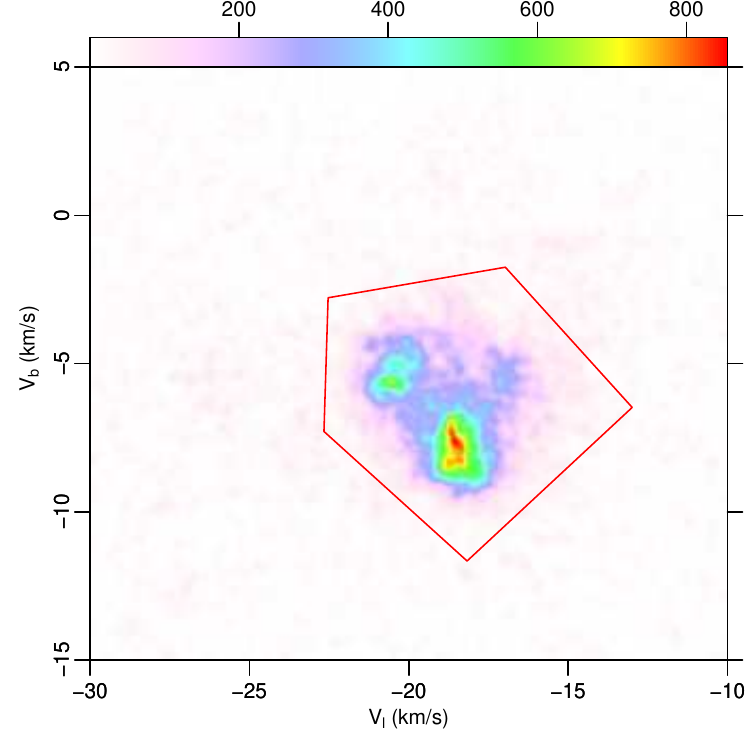}}
\caption{
{
Transverse-velocity plot $(V_l, V_b)$, for PMS and Upper-MS stars in the
diffuse component.  The red box is the same as in Fig.~\ref{vt-all}.
}
\label{diffuse-vt}}
\end{figure}

\begin{figure}
\resizebox{\hsize}{!}{
\includegraphics[]{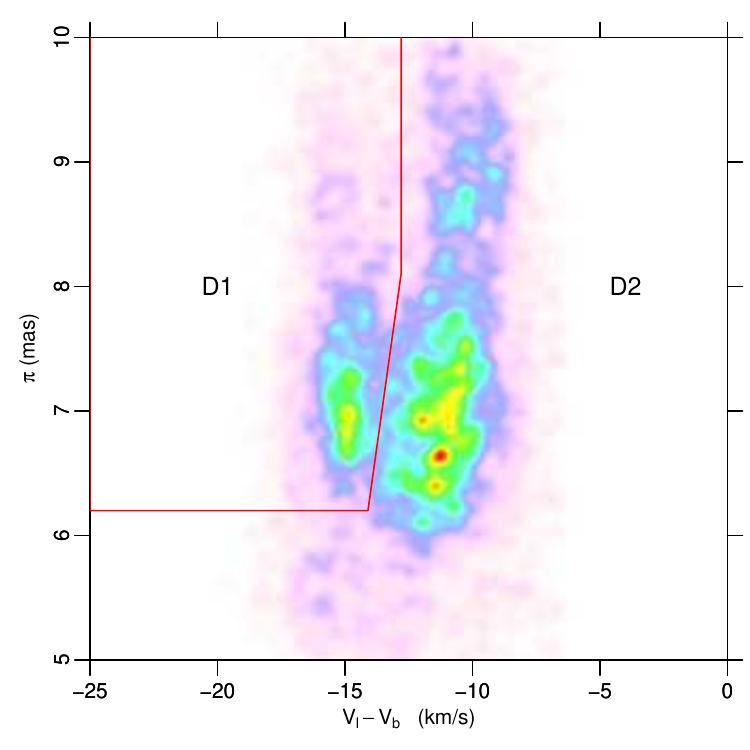}}
\caption{
{
Plot of parallax $\pi$ vs.\ transverse-velocity difference $V_l-V_b$.
The solid red line separates diffuse components D1 and D2, as labeled.
}
\label{diffuse-vt-plx}}
\end{figure}

\begin{figure*}
\resizebox{\hsize}{!}{
\includegraphics[]{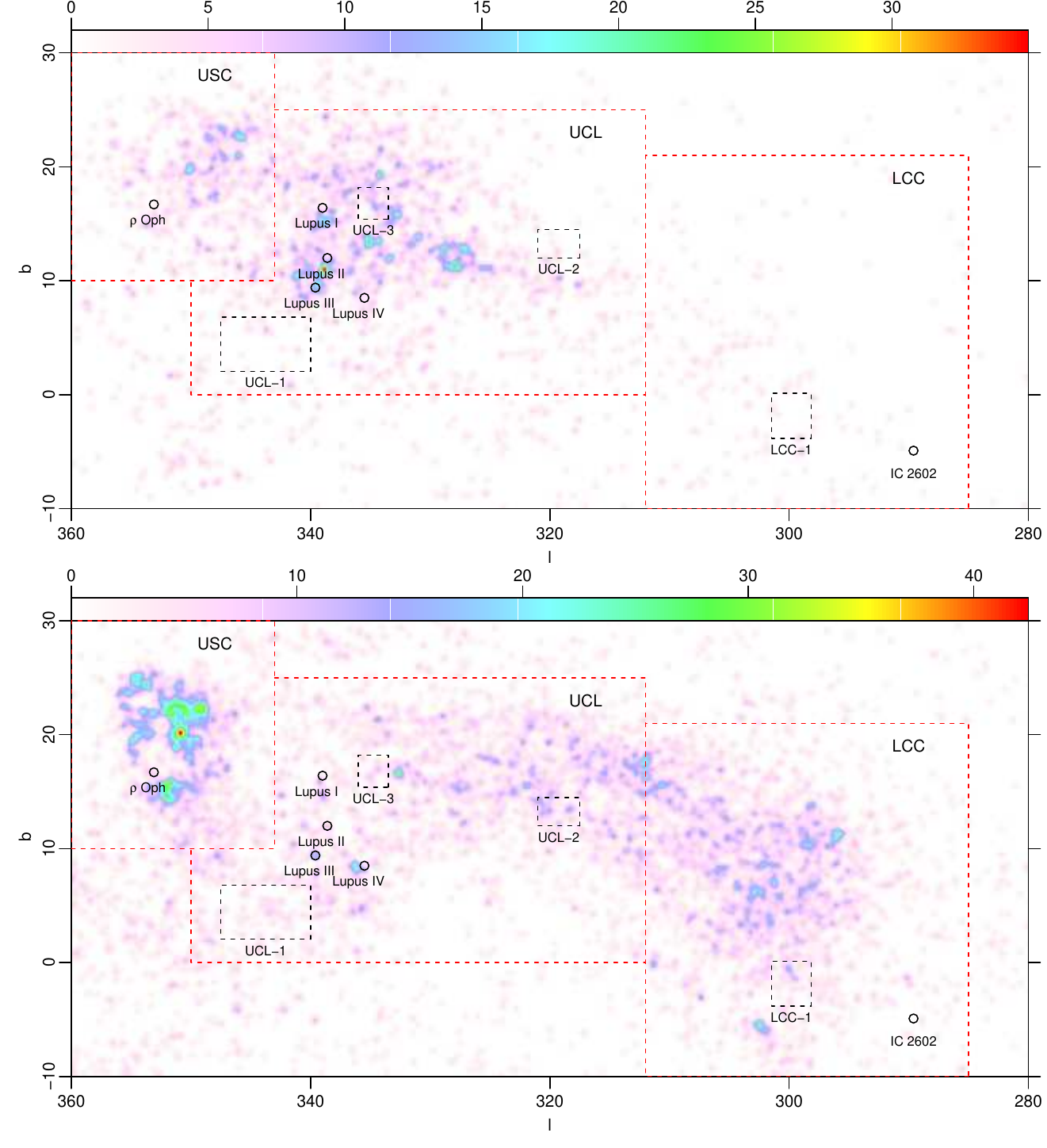}}
\caption{
{
Spatial maps of diffuse components D1 (upper panel), and D2 (lower panel).
}
\label{diffuse-space}}
\end{figure*}

The number of Sco~OB2 members found to belong to compact groups in the
previous subsection is 2088 (1913 PMS and 175 Upper-MS stars). This is
only 14.5\% of the total number of PM$+$CAMD selected members: the bulk
of association members are found in the diffuse population. In this
section we use methods presented in Sect.~\ref{space} to study the
properties of kinematical groups among this large diffuse population.
We have tried several methods to identify the best candidate samples for
being a kinematical group. We here start from the distribution on the VT
plane of diffuse members, shown in Fig.~\ref{diffuse-vt}. Still after
removal of the compact populations, a complex multi-peaked structure is
observable in the VT plane. We might ask if projection effects can be
held responsible for such a structure (as depth effect were demonstrated
to be in the case of some of the structures on the PM plane), but this is
unlikely the case, for at least two reasons: 1) the association is much more
elongated along $l$ than along $b$, yet the two main peaks in the VT plane 
are no better separated along $V_l$ than along $V_b$; and 2) the spatial
distribution of the (dominant) diffuse component in Sco~OB2 is too
smooth to produce two distinct peaks such as those in Fig.~\ref{diffuse-vt}.
Multiple peaks on the VT plane therefore correspond likely to different
kinematical groups. This is further confirmed by the density plot shown
in Fig.\ref{diffuse-vt-plx}, showing $\pi$ vs.\ the velocity difference
$V_l-V_b$ (i.e., the projection of the velocity vector along the
direction which maximizes the group differences, times $\sqrt{2}$).
The richest group (labeled D2 in the figure) spans a much larger
parallax range, and probably consists of multiple sub-populations; the
less rich group (D1) spans a smaller $\pi$ range, and is more likely to
be a single population.

All diffuse populations defined in this section are required to be
VT-selected members, in order to minimize the number of possible
interlopers. As a consequence, there will be a number of PM-selected
members falling neither in compact nor in diffuse populations, but which
are still kept in order to avoid excessive incompleteness, as explained
above. The population which each star is assigned to is also reported in
Tables~\ref{table-pms} and~\ref{table-upp-ms}.

The spatial distributions of populations D1 and D2 are shown in
Fig.~\ref{diffuse-space}. D1 is mostly found in UCL, and nearly absent in LCC,
while D2 populates all three of USC, UCL, and LCC. The density of D2 is
highest in USC, and there is a hint of a density gap between USC and the
rest of the D2 population. On the other hand, D2 appears to extend
beyond the commonly adopted boundaries of Sco~OB2, for example in the
sky region $350<l<360$ and $0<b<10$ (hosting the cloud Barnard~59 and
the Pipe Nebula).

\begin{figure}
\resizebox{\hsize}{!}{
\includegraphics[angle=90]{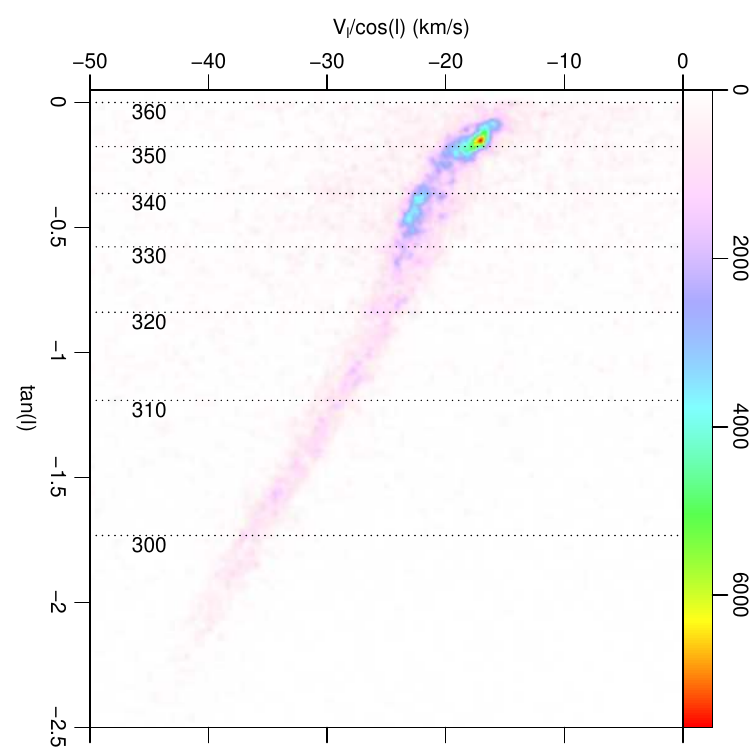}}
\caption{
{
Plot of $\digamma = V_l \, /\cos l$ vs.\ $\xi = \tan l$ for all diffuse
components. Vertical dotted lines are plotted at constant $l$, spaced by
10 degrees.
}
\label{diffuse-xidig}}
\end{figure}

Adopting the formalism developed in Sect.~\ref{space}, we show the
$(\digamma, \xi)$ diagram for the whole diffuse population
in Fig.~\ref{diffuse-xidig}. While for $l<320$ a linear dependence of
$\digamma$ on $\xi$ (and therefore a single dynamical population)
is consistent with the data, this is not the case for $l>320$.
This confirms that multiple dynamical populations do exist, regardless
of projection effects. The same diagram, for the D1 component as defined
above, is shown in Fig.~\ref{diffuse-xidig-d1}. In contrast to
Fig.~\ref{diffuse-xidig}, here a linear relation between $\digamma$ and $\xi$
is observed, and the relative best-fit line is shown. Also shown are the
positions of the compact populations, computed using the respective median
transverse velocities and positions. The UCL-1 and UCL-2 groups have a
dynamics which is very close to the D1 diffuse component, while UCL-3 is
only slightly discrepant; the dynamics of all other compact groups is much less
consistent with that of population D1. As explained in
Sect.~\ref{space}, intercept and (negative) slope of the best-fit line to D1
provide space velocities $V_0=-17.86$~km/s and $U_0=-10.11$~km/s
(Eq.~\ref{eq16}). Inserting these
constants in Eq.~\ref{eq17} we obtain (star-by-star) estimates of $W$
for population D1: this will be confirmed to consist of a well-defined
kinematical population if the inferred values of $W$ fall into a narrow
range.  This is indeed the case, as demostrated by the histogram of
derived $W$ values in Fig.~\ref{diffuse-whist-d1}, having a peak
at $W_0=-5.97$~km/s, and a std.dev.\ of 1.07~km/s. The peak width is well
consistent with the velocity spread commonly found in young clusters
(1-2~km/s) and we conclude that D1 is really a well-defined kinematical
subpopulation according to our method~A.

We have checked this result using our method~B: to do this, we use
Eqs.~\ref{eq18} to~\ref{eq20}, generating for each star all possible
positions $(U,V,W)$ in velocity space for $v_R$ in the range
[-25,25]~km/s, in steps of 0.5~km/s. Figure~\ref{methB-d1} shows the
outcome of this procedure, as density of datapoints projected to the
$(V,U)$ and $(V,W)$ planes, respectively in left and right panels.
In both panels, the density peak corresponds well to the values found
using method~A (small circles in the figure). On the $(V,W)$ plane,
method~B shows some substructures in the density peak; however, they are
not sufficiently separated to permit a more refined dynamical
classification.

\begin{figure}
\resizebox{\hsize}{!}{
\includegraphics[]{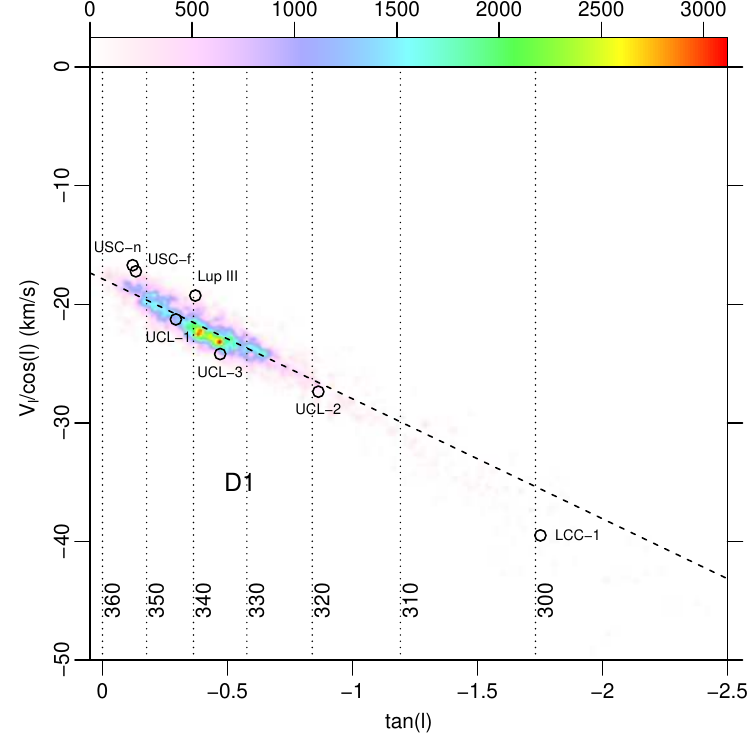}}
\caption{
{
$(\digamma, \xi)$ plot for the D1 component. Vertical dotted lines as in
Fig.~\ref{diffuse-xidig}. The dashed line is a linear best fit to the D1
datapoints. The labeled circles indicate the median $(\digamma, \xi)$
values of the compact Sco~OB2 populations.
}
\label{diffuse-xidig-d1}}
\end{figure}

\begin{figure}
\resizebox{\hsize}{!}{
\includegraphics[angle=90]{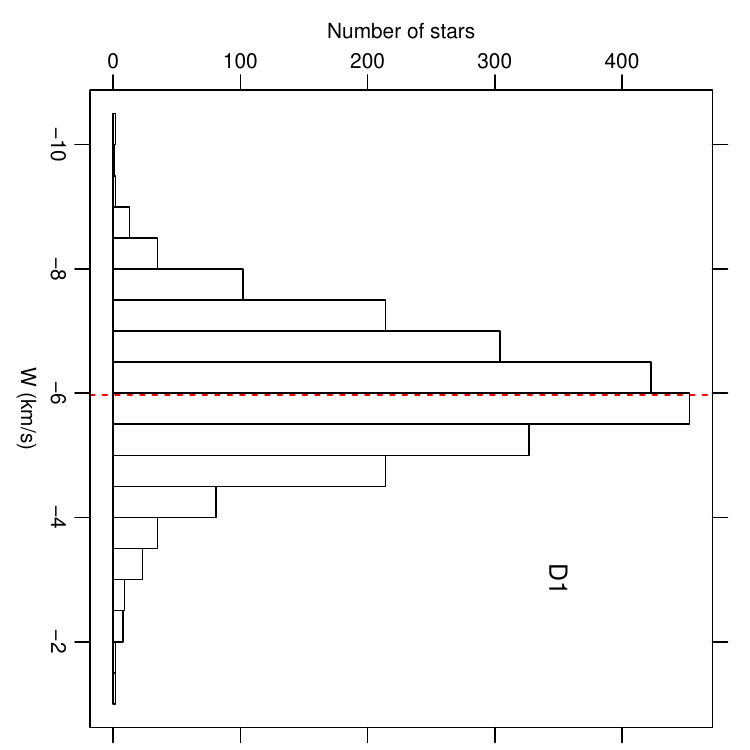}}
\caption{
{
Histogram of inferred $W$ for the D1 population. The vertical red dashed
line indicates the median $W$ (-5.97 km/s).
}
\label{diffuse-whist-d1}}
\end{figure}

\begin{figure*}
\resizebox{\hsize}{!}{
\includegraphics[]{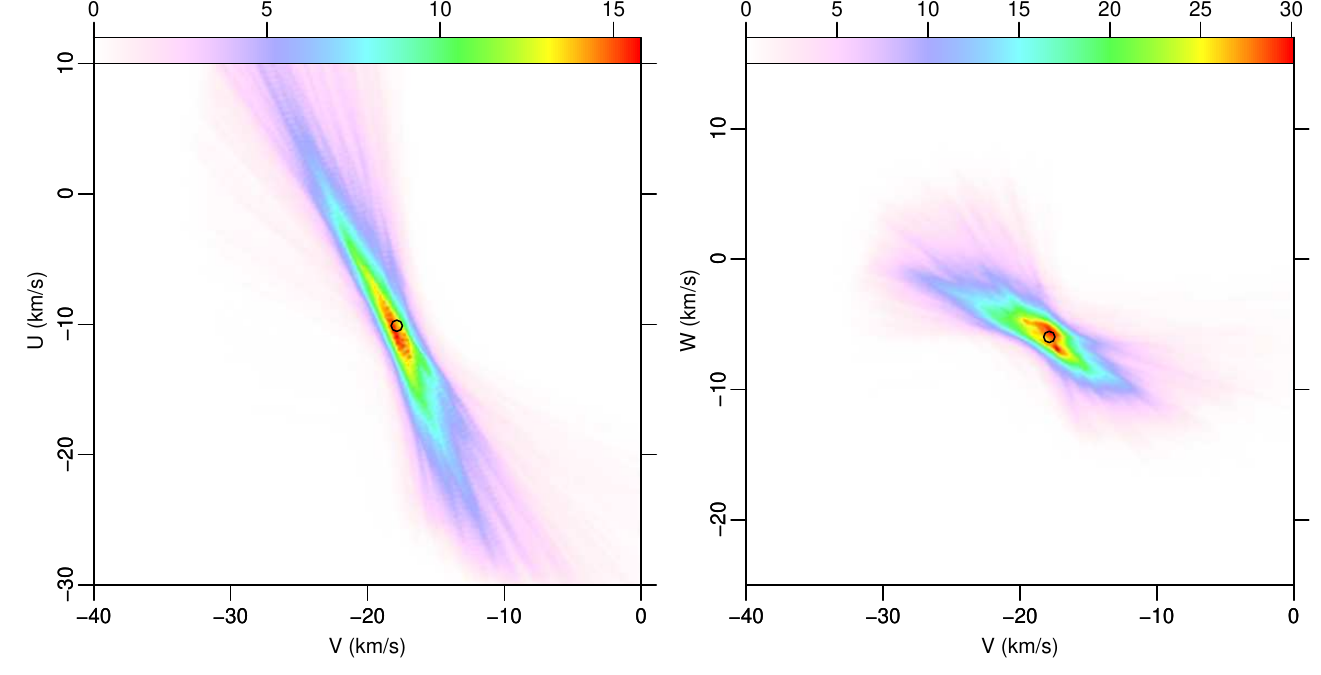}}
\caption{
{
Results of our method~B, for the D1 population.
Left panel: $(U,V)$ density plot.
Right panel: $(W,V)$ density plot.
In each panel, the circle indicates the result of method~A.
}
\label{methB-d1}}
\end{figure*}

An analogous procedure was followed for population D2:
Fig.~\ref{diffuse-xidig-d2} is the corresponding $(\digamma, \xi)$
diagram. We have excluded from D2 the USC region, since, as noted above,
the spatial distribution of its diffuse component does not appear as a
simple extension of D2 into the USC region. Even after removing the USC
region, Fig.~\ref{diffuse-xidig-d2} shows that D2 deviates from a simple
straight line at $\tan l \sim -0.5$. For $\tan l >-0.45$ the diagram
appearance suggest the coexistence of two populations, labeled in the
figure as D2a and D2b, separated by the blue dotted line, and
following distinct linear loci (dashed lines). The D2a population
overlaps the $\tan l$ range of D1 (Fig.~\ref{diffuse-xidig-d1}), but is
a dynamically different population: the best-fit line for D1 is also
shown in Fig.~\ref{diffuse-xidig-d2} (red dotted line) for comparison.
D2a is instead kinematically consistent with compact groups USC-near and
-far. D2b is consistent with Lupus~III, and slightly less with LCC-1,
while it is obviously inconsistent with UCL-1 and UCL-3.
For D2a we obtain best-fit values of $V_0=-14.94$~km/s and
$U_0=-16.2$~km/s. Again, we infer $W$ values for its member stars, but
this time we obtain a more complex $W$ histogram, as shown in
Fig.~\ref{diffuse-whist-d2.1}: the $W$ distribution is obviously
doubly-peaked, with maxima near $W=-5.2$~km/s and $W=-8.75$~km/s. This
is best interpreted as two populations, with nearly undistinguishable
$U$ and $V$ components, but well-separated $W$ components. An
approximate boundary in $W$ between the two sub-population is adopted at
$W=-6.75$ (dotted red line in Fig.~\ref{diffuse-whist-d2.1})
We turn to
method~B for a check, whose results are shown in Fig.~\ref{methB-d2.1},
the analogous of Fig.~\ref{methB-d1}. Here again, the $(V,U)$ plane
shows a single fairly well defined peak, coincident with that found
using method~A; in the $(V,W)$ plane, instead, two distinct linear streams
are evident, which are well consistent with the two D2a sub-populations
found by method~A. Again from this diagram, the sub-population
corresponding to peak $W=-5.2$~km/s is less rich than the other, and a
complete separation between the two does not seem possible.

\begin{figure}
\resizebox{\hsize}{!}{
\includegraphics[]{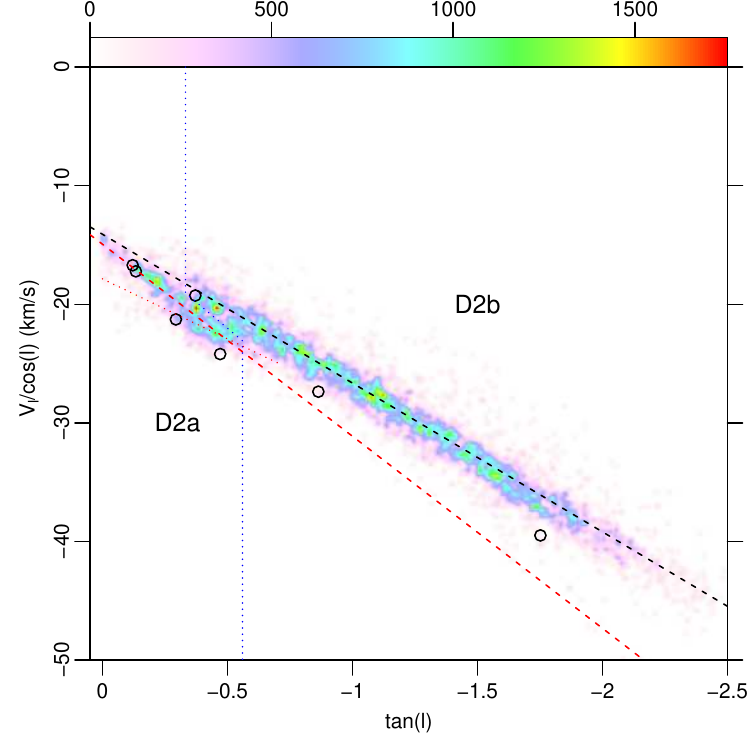}}
\caption{
{
$(\digamma, \xi)$ plot for the D2 component, but excluding the USC region.
Vertical dotted lines as in Fig.~\ref{diffuse-xidig}.
The dotted blue line separates diffuse sub-populations D2a and D2b,
whose best-fit lines are shown as dashed red and black lines, respectively.
The dotted red line is the D1 best-fit from Fig.~\ref{diffuse-xidig-d1},
as a reference.
The small circles are as in Fig.~\ref{diffuse-xidig-d1}.
}
\label{diffuse-xidig-d2}}
\end{figure}

\begin{figure}
\resizebox{\hsize}{!}{
\includegraphics[angle=90]{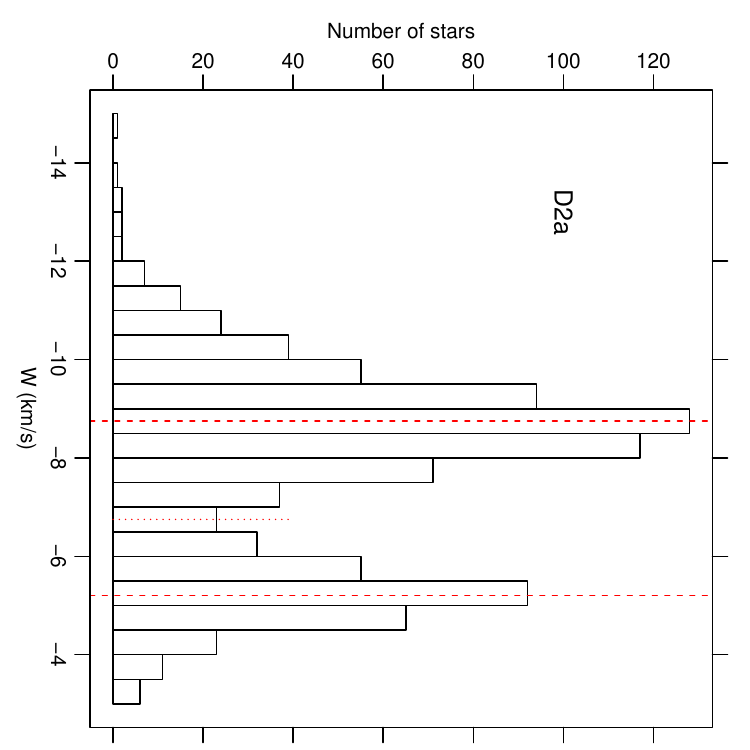}}
\caption{
{
Histogram of inferred $W$ for the D2a population. The vertical red dashed
lines are estimates of the median $W$ for the two peaks.
The boundary between the two D2a sub-populations is assumed at $W=-6.75$
(dotted red segment).
}
\label{diffuse-whist-d2.1}}
\end{figure}

\begin{figure*}
\resizebox{\hsize}{!}{
\includegraphics[]{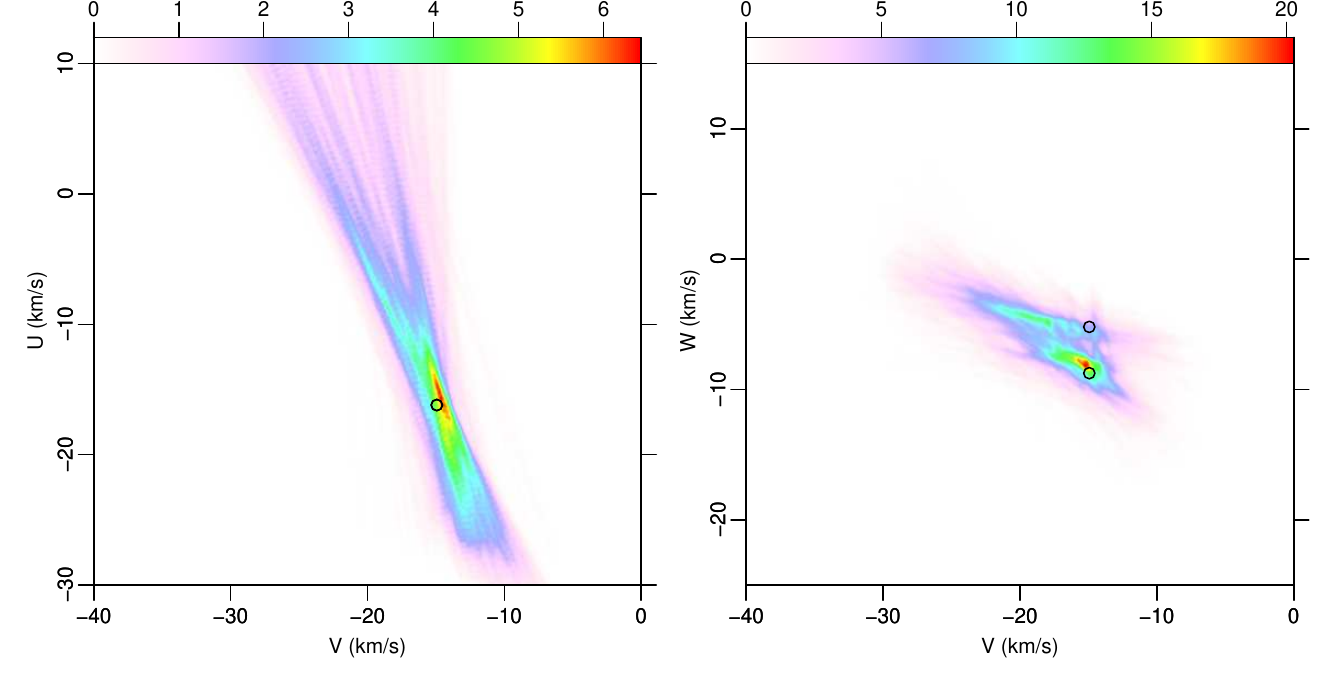}}
\caption{
{
Results of our method~B, for the D2a population.
Left panel: $(U,V)$ density plot.
Right panel: $(W,V)$ density plot.
In each panel, circles indicate the results of method~A.
}
\label{methB-d2.1}}
\end{figure*}

\begin{figure}
\resizebox{\hsize}{!}{
\includegraphics[angle=90]{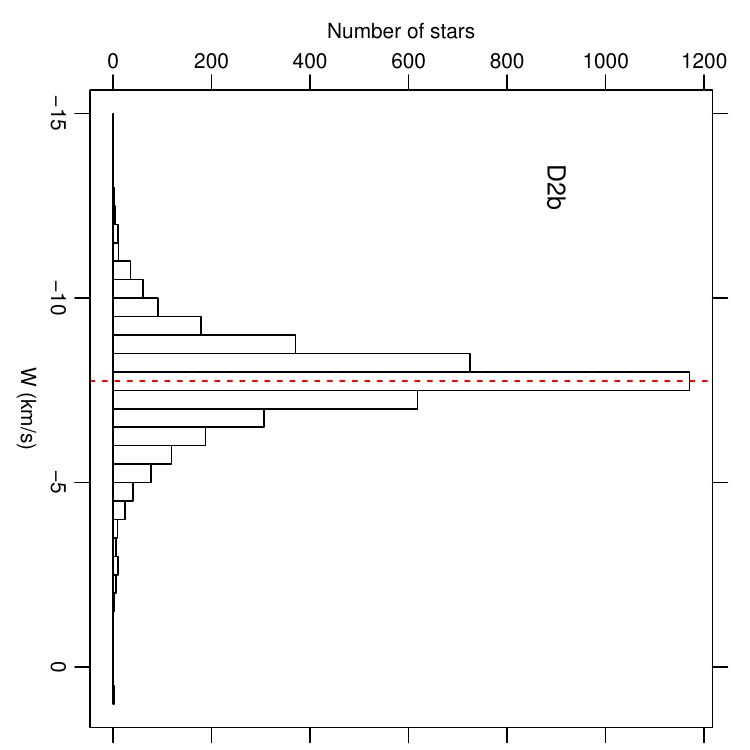}}
\caption{
{
Analogous of Fig.~\ref{diffuse-whist-d1}, for the D2b population.
}
\label{diffuse-whist-d2.2}}
\end{figure}

\begin{figure*}
\resizebox{\hsize}{!}{
\includegraphics[]{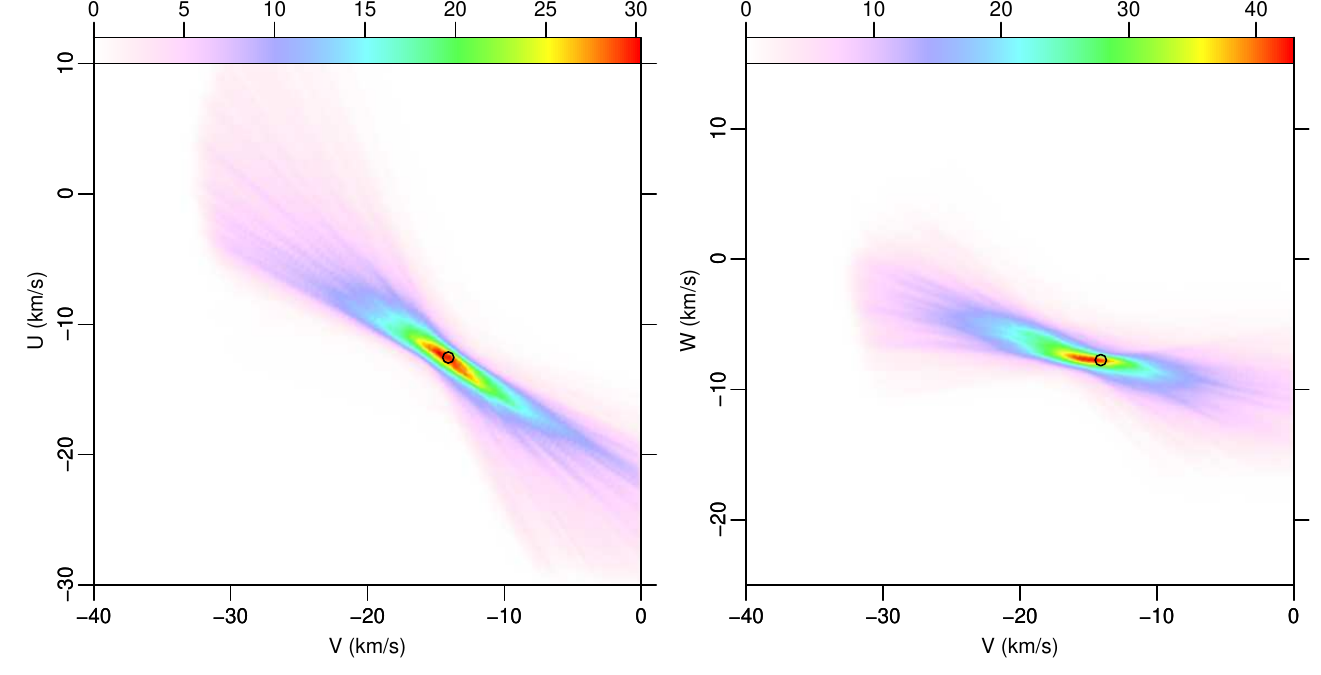}}
\caption{
{
Results of our method~B, for the D2b population.
Left panel: $(U,V)$ density plot.
Right panel: $(W,V)$ density plot.
In each panel, the circle indicates the result of method~A.
}
\label{methB-d2.2}}
\end{figure*}

Finally, diagrams illustrating results from methods ~A and~B for D2b are
shown in Figs.~\ref{diffuse-whist-d2.2} and~\ref{methB-d2.2}. In this
case we obtain $V_0=-14.09$~km/s and $U_0=-12.55$~km/s, and the $W$
histogram is well-behaved, with a peak (median $W$) at $W=-7.75$ and
std.dev.\ of 1.2~km/s. Correspondingly, method~B also indicates a single
population (Fig.~\ref{methB-d2.2}), with space velocities in agreement
with method~A.

Total number of PMS (Upper-MS) members for D1, D2a and D2b populations
are respectively 2105 (353), 757 (145), and 3562 (640).
The existence of discrete diffuse populations, each with its own
well-characterized kinematical properties, as found here, is in contrast
with the continuous dependence of $(U,V,W)$ on $l$, as in Rizzuto \e
(2011). In hindsight, in Figure~2$a$ from Rizzuto \e, at least three layers of
constant $U$ were already noticeable. The Gaia DR2 data are much more
precise than those available to those authors, so that a continuous $U$
distribution across the entire Sco~OB2 can be ruled out.

\subsubsection{Diffuse population in Upper Sco}
\label{diff-usc}

As is clear from Sect.~\ref{usc-compact}, the USC region hosts
populations with a complex structure, both spatially and kinematically.
Once we remove the compact populations found in Sect.~\ref{usc-compact},
the remaining diffuse USC population (named USC-D2, since it is a D2
sub-population) distributes on the PM plane as
shown in Fig.~\ref{usc-diffuse-pm}. Again, projection effects over such
a limited sky region are negligible. A comparison between
Fig.~\ref{usc-diffuse-pm} and Fig.~\ref{usc-compact-pm} shows that the
dominant peak in this latter (USC-near, at $\mu_b<-10$~mas/yr) has no
correspondence among stars in the diffuse population
(Fig.~\ref{usc-diffuse-pm}). Conversely, only secondary peaks in
Fig.~\ref{usc-diffuse-pm} (at $\mu_l>-24$~mas/yr) correspond to stars in
USC-far in Fig.~\ref{usc-compact-pm}. Therefore, compact and diffuse
population in USC have largely non-overlapping kinematical properties,
except perhaps for the USC-far group.
Nevertheless, Fig.~\ref{usc-diffuse-pm} suggests the existence of
kinematical substructures (or multiple subpopulations) among the diffuse
USC-D2 population, analogous though not identical to compact USC
populations.

\begin{figure}
\resizebox{\hsize}{!}{
\includegraphics[]{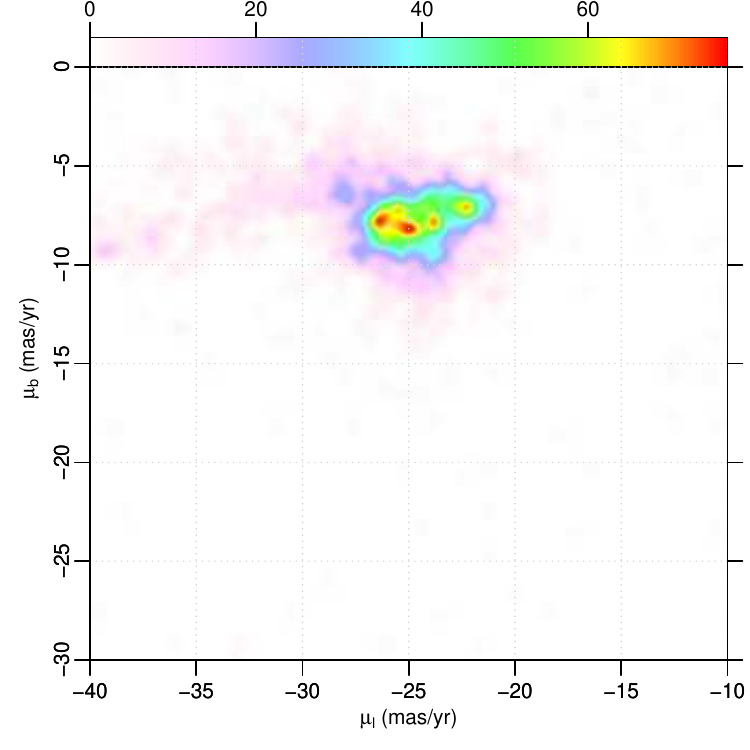}}
\caption{
{
PM diagram of the diffuse population in USC.
}
\label{usc-diffuse-pm}}
\end{figure}

\begin{figure}
\resizebox{\hsize}{!}{
\includegraphics[]{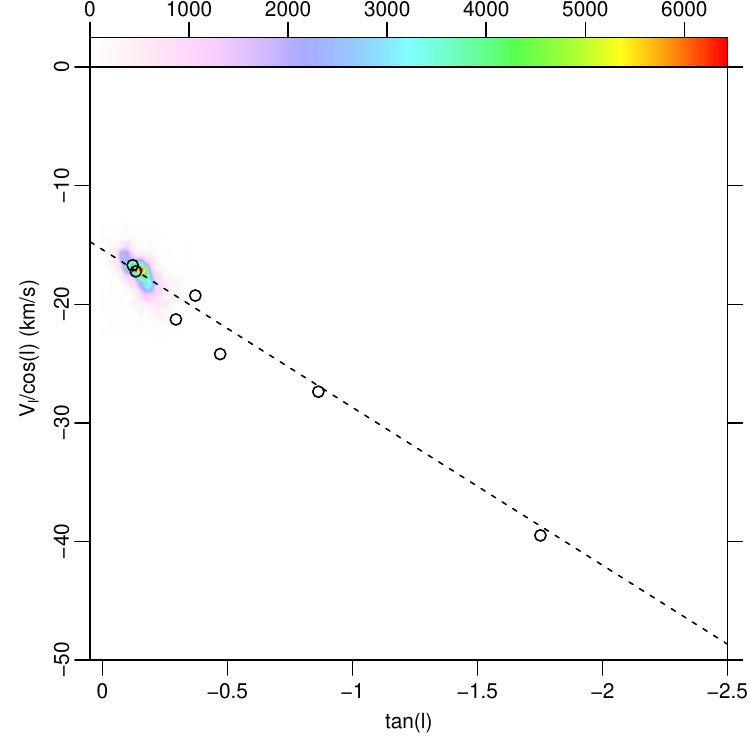}}
\caption{
{
$(\digamma, \xi)$ plot for the USC-D2 population.
The dashed line is the best fit.
The small circles are as in Fig.~\ref{diffuse-xidig-d1}.
}
\label{diffuse-xidig-d2.3}}
\end{figure}

\begin{figure}
\resizebox{\hsize}{!}{
\includegraphics[]{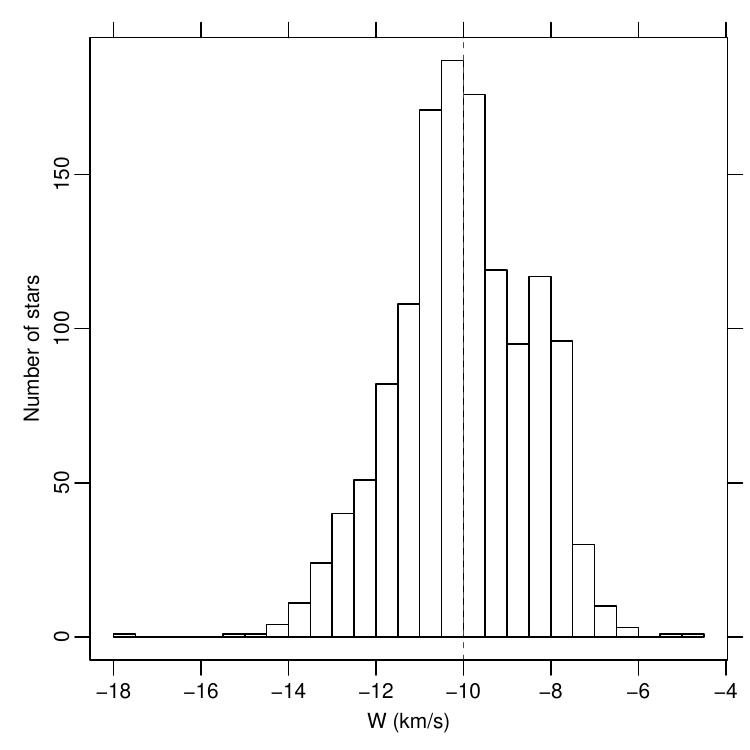}}
\caption{
{
Analogous of Fig.~\ref{diffuse-whist-d1}, for the USC-D2 population.
}
\label{diffuse-whist-d2.3}}
\end{figure}

Since projection effects across USC are small, our methods~A and~B are
unlikely to provide significant indications on the space velocity vector
of the USC-D2 population. We have nevertheless attempted this
route as above, with the results shown in Figs.~\ref{diffuse-xidig-d2.3}
and~\ref{diffuse-whist-d2.3} for method~A, and Fig.~\ref{methB-d2.3}
for method~B.

Fig.~\ref{diffuse-xidig-d2.3} show that the $(\xi, \digamma)$ diagram
for USC-D2 is somewhat wiggly, and a linear fit not very significant.
The inferred $W$ histogram of Fig.~\ref{diffuse-whist-d2.3} is, not
unexpectedly, less regular than its analogues for D1 or D2b; its
std.dev.\ is larger (1.54~km/s), although still plausible.
Method~B confirms that results from method~A are not well constrained:
the locus of maximum density in the $(V,U)$ plane (left panel in
Fig.~\ref{methB-d2.3}) is very elongated (degenerate parameters);
moreover, the density distribution in the $(V,W)$ plane possibly suggests 
two loci for the maxima, which are however very hard to disentangle, and
only in rough agreement with results from method~A. This means that the
structures on the PM plane of Fig.~\ref{usc-diffuse-pm} for the USC-D2
population are likely to be real, but there is no simple, non-arbitrary
way to separate the subpopulations giving rise to them.
Overall, USC-D2 comprises 1210 PMS and 119 Upper-MS members.

\begin{figure*}
\resizebox{\hsize}{!}{
\includegraphics[]{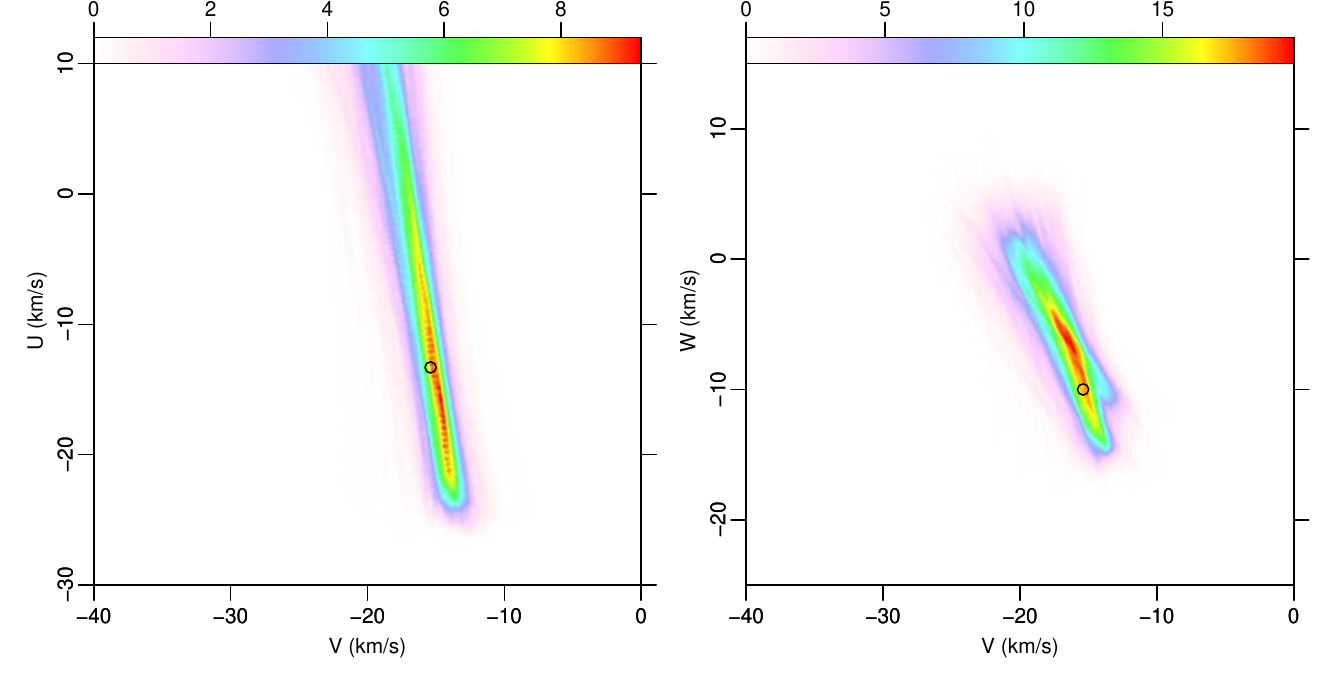}}
\caption{
{
Results of our method~B, for the USC-D2 population.
Left panel: $(U,V)$ density plot.
Right panel: $(W,V)$ density plot.
In each panel, the circle indicates the result of method~A.
}
\label{methB-d2.3}}
\end{figure*}

\begin{figure*}
\resizebox{\hsize}{!}{
\includegraphics[]{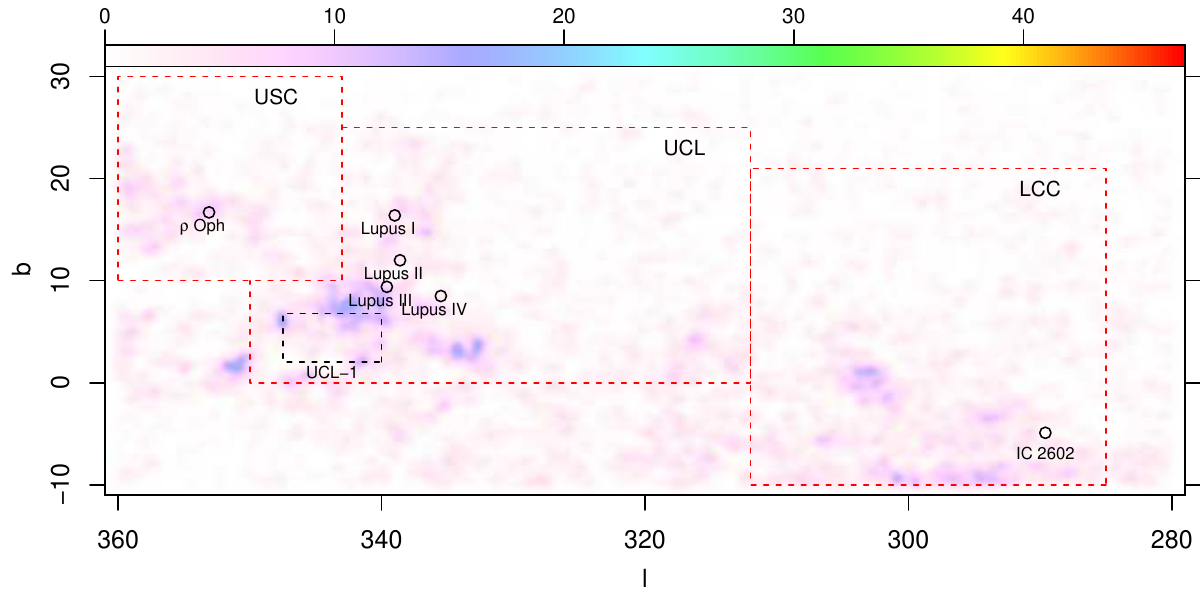}}
\caption{
{
Spatial map of PM-selected PMS candidate members having parallaxes
$3.3 < \pi < 5$ (distance between 200-300~pc).
The color scale is the same as in Fig.~\ref{spatial-map-all-satur} 
for an immediate comparison.
}
\label{spatial-map-beyond}}
\end{figure*}

\begin{figure}
\resizebox{\hsize}{!}{
\includegraphics[]{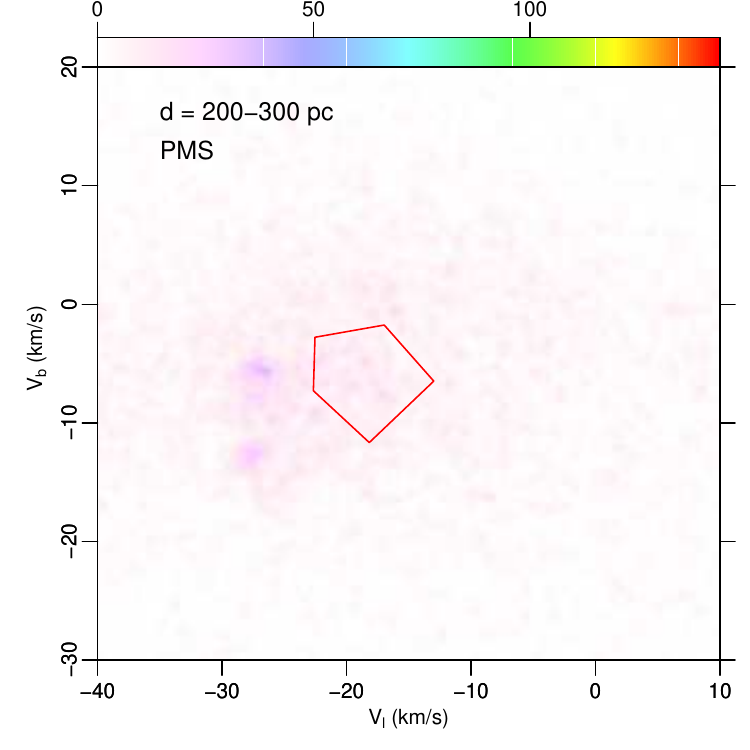}}
\caption{
{
Transverse-velocity $(V_l, V_b)$ plot of PMS stars between 200-300~pc.
The red box is the same as in Fig.~\ref{vt-all}.
}
\label{vt-pms-beyond}}
\end{figure}

\subsection{Stars beyond 200 pc}
\label{beyond}

In Sect.~\ref{obs} we mentioned that $\sim 8$\% of Sco~OB2 members from
de~Zeeuw (1999) have Gaia parallaxes indicating distances larger than
200~pc. However, we also found in Sect.~\ref{pm} that only less than 1\%
of low-mass members from PM16 lie farther than 200~pc.
Here we try to estimate if any significant part of Sco~OB2 lies farther
than this distance, which was assumed as the effective far boundary of
the association in all previous sections. Therefore, we selected all
Gaia sources with $3.3 < \pi < 5$ (distance range 200-300~pc)
and $\pi/ \Delta \pi >10$ as before,
with PM-plane selection and falling inside the PMS locus on the
CAMD\footnote{We checked that the larger distance of the 200-300~pc
sample implies a faint limit around $M_G \sim 13$ on the corresponding
CAMD. In Fig.~\ref{camd-pms-define} only 70 PMS members ($\sim 0.6$\% of
all PMS members) fall below that limit. Accordingly, we do not expect
that the comparison between samples within and beyond 200~pc is vitiated
by their different limiting $M_G$s.}.
For stars up to 200~pc, the same selection criteria resulted in the
spatial map shown in Fig.~\ref{spatial-map-all-satur}. For the 200-300
distance range, the resulting sky distribution is instead shown in
Fig.~\ref{spatial-map-beyond}, displayed with the same color scale as
Fig.~\ref{spatial-map-all-satur}. The density of these (candidate
PMS) stars is not uniform, with 3-4 probable concentrations (mainly near
Lupus), but does not follow the spatial pattern found for Sco~OB2
members. One of the most significant overdensities, as also part of the
diffuse component, lie even outside the
conventional association boundary. Average and peak densities are much lower 
than the corresponding values in Sco~OB2. All these characteristrics
argue against these stars being association members.

A further indication in this sense is provided by the VT diagram for the
same stars (Fig.~\ref{vt-pms-beyond}). The VT diagram is preferred to the PM
diagram because is compensates differences in depth, which are there
by definition. The comparison between this figure and Fig.~\ref{vt-pms}
shows that there is no concentration of PMS candidates between
200-300~pc inside the red polygon used to select Sco~OB2 members; on the
other hand, two distinct density peaks in the VT plane are found at
larger negative $V_l$ values, outside the red polygon. Most of the
diffuse population between 200-300~pc is widely spread over the VT plane.
If the color scale in Fig.~\ref{vt-pms-beyond} were the same as
in Fig.~\ref{vt-all}, the diagram would appear as almost pure white.
We conclude that there is no significant presence of Sco~OB2
members beyond 200~pc. Therefore, de~Zeeuw members beyond that distance
are either non-members, or perhaps runaway members.
The first possibility is entirely plausible, as in Table~C1 from
de~Zeeuw \e (1999) 8\% of the Sco~OB2 members have membership
probabilities $P<70$\%.
The scarcity of PM16 members beyond 200~pc agrees well with our results.
}

\section{Stellar ages}
\label{ages}

\begin{figure*}
\resizebox{\hsize}{!}{
\includegraphics[angle=0]{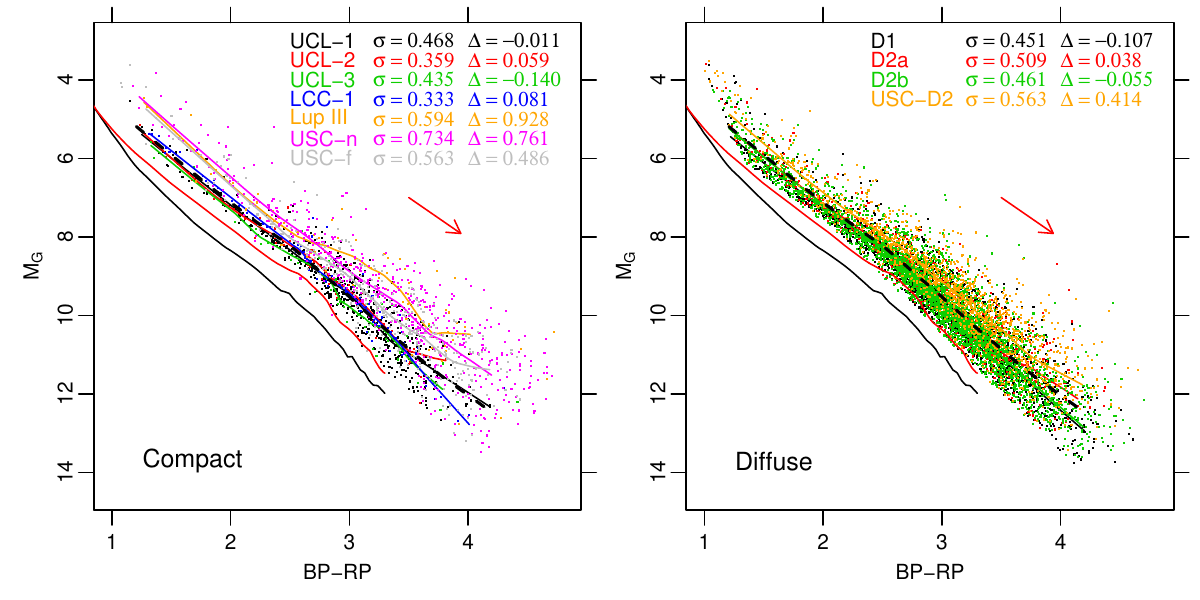}}
\caption{
Left panel: CAMD of stars in the { compact populations}, with
colors indicating membership.
Right panel: CAMD of stars in the { diffuse populations}.
The thin black (red) solid line is a best fit to the Pleiades (IC~2602)
sequence, from Gaia DR2 photometry.
For each { population}, $\sigma$ (mag) is the standard deviation of
datapoints around the respective best-fit (colored solid lines), while $\Delta$
(mag) is the mean difference with respect to the overall best-fit (thick
black dashed line).
{ The reddening vector is shown with red arrows.}
\label{camd-diffuse-compact}}
\end{figure*}

An important piece of information on the star-formation history in
Sco~OB2 is contained in the CAMD already presented in
Fig.~\ref{camd-pms-define}. In particular, the obvious gap between the
MS and PMS loci shows that star formation in Sco~OB2 only started less
than 30-40~Myr (at most), and was essentially absent at earlier times,
that is, before the birth of IC~2602, whose sequence is shown in
Fig.~\ref{camd-pms-define}.
If continuous star formation were present, one would expect in the CAMD a star
density inversely proportional to the speed at which a star crosses a
given part of the diagram: therefore, datapoints should cluster near the
positions corresponding to the latest PMS stages, and have the least
density high up on the Hayashi track, that is the opposite of what we
observe.
The pre-selection of sources shown in our CAMD was made only from the
(wide) region of the PM plane from Fig.~\ref{pm-nofilt}. Strictly
speaking, we can only rule out that the parent molecular cloud of today's
Sco~OB2, whose dynamical signature is seen in Fig.~\ref{pm-nofilt},
formed stars earlier than 30~Myr ago: there might be many stars at
intermediate ages (e.g., 100-300~Myr) in the same space region, but
having different kinematical properties.
Alternatively, the Sco~OB2 parent cloud might have given birth to an
earlier generation of stars, but their dynamical signatures were
completely erased on timescales of less than 30-40~Myr, through
dynamical interactions with field stars. This latter scenario seems
however less likely, if we consider the rather narrow width of the PMS
locus in Fig.~\ref{camd-pms-define}, and the well-defined spatial
boundaries of the diffuse Sco~OB2 population found in
Fig.~\ref{spatial-map-all-satur}.

We have examined the individual CAMDs of { all kinematical
populations defined in Sect.~\ref{spatial}. Members of IC~2602 are 
not considered here.
We differentiate between the spatially compact and diffuse populations.}
Since stellar aggregates are known to disperse with time
(on average), but never to condense out of a dispersed population,
the diffuse Sco~OB2 populations must be on average older than the compact
populations. This holds irrespective of our ability to trace back star
positions to their birthplace, from accurate measurements of individual
stellar motions.

The CAMDs of Sco~OB2 stars from compact { populations} are shown in
Fig.~\ref{camd-diffuse-compact} (left panel). For each group, a
non-parametric fit (lowess) was made. The same was made for the Gaia DR2
data of the Pleiades and IC~2602 (solid lines) and for the total PMS
Sco~OB2 population (dashed black line). The standard deviations $\sigma (M_G)$
(mag) around the individual best-fits, and the average difference $\Delta
(M_G)$ (mag, positive upwards) with respect to the total-population best-fit
were computed for each subgroup, as shown in the figure legend.
The same procedure was made for the diffuse { populations}, and the result
shown in the right panel of the same figure.
{
Even though we did not correct for reddening, it is clear from the
figure that it cannot be responsible for the observed spread of datapoints.
}

{
The comparison among all populations is easier if we consider the
diagram in Fig.~\ref{age-delta-sigma}, which summarizes all $\sigma$ and
$\Delta$ values from the previous CAMDs.
The youngest compact populations are (in order of increasing age)
Lup-III, USC-near and USC-far. The youngest diffuse population is USC-D2
which has both $\sigma$ and $\Delta$ very close to USC-far: these two
populations are closely related. The remaining diffuse populations are
definitely older, with ages not significantly different from the compact
groups UCL-1, UCL-2, UCL-3 and LCC-1.
D1 and D2b populations constitute the bulk of UCL and LCC members (without
one-to-one correspondence to these spatial regions, however), and their
indistiguishable ages match well with results from Mamajek \e (2002) and
PM16.

\begin{figure}
\resizebox{\hsize}{!}{
\includegraphics[]{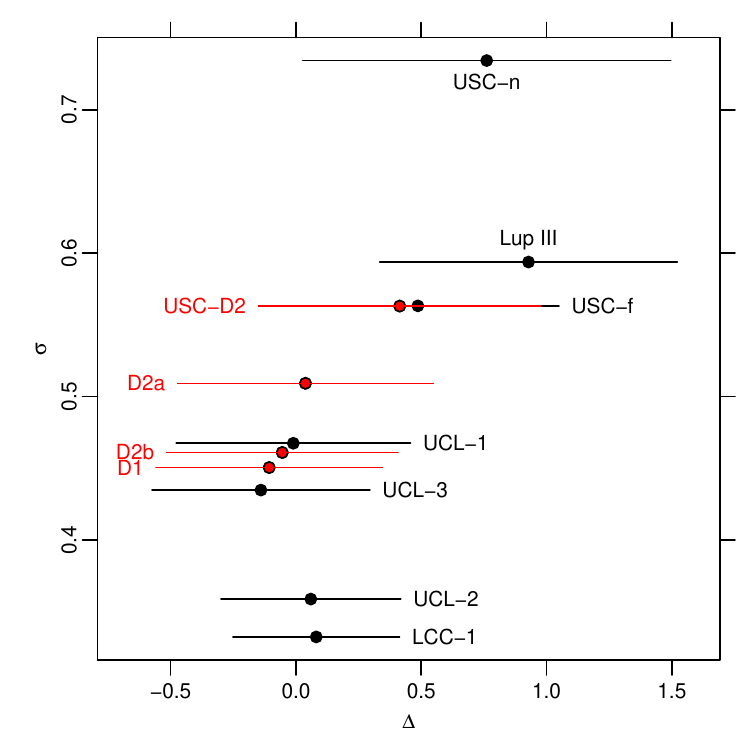}}
\caption{
{
Magnitude difference spread $\sigma$ vs.\ average difference $\Delta$
(mags). Error bars on $\Delta$ are equal to $\sigma$. In red are diffuse
populations.
}
\label{age-delta-sigma}}
\end{figure}

The fact that the three youngest groups are compact ones
}
agrees with our above arguments that the diffuse
Sco~OB2 populations are on average older than the compact ones.
However, it is not entirely satisfactory
{
that a non-negligible fraction of stars in USC-near and -far are found
at old apparent ages (small $\Delta$), overlapping the bulk of stars
in the D2b diffuse population. The latter is mostly located in LCC.
}
As extensively discussed in the
literature reviewed by Preibisch and Mamajek (2008), USC is thought to
be the youngest part of Sco~OB2 (only 22 stars older than 5~Myr
were found in the $\rho$~Oph cluster by Pillitteri \e 2016);
moreover, star formation in USC has
been argued to have been triggered by events (e.g., supernova
explosions) in the neighboring UCL region, itself triggered from LCC
(the oldest part of Sco~OB2). The presence in USC of stars as old as
those in LCC does not fit in this picture. The only possibility to
reconcile the commonly accepted sequence of star-formation events across
Sco~OB2 with the { incongruence described above} is that
individual star positions in the CAMD do not reflect (only) stellar
ages. This was already proposed by Baraffe \e (2012, 2017),
and Dunham \& Vorobyov (2012); see also Jeffries (2012). According to
these models, the position of a star in the PMS part of the CAMD (or any
equivalent of the temperature-luminosity diagram) would depend not only
on its mass and age, but also on its past accretion history during the
protostellar phase, which may
be different from star to star. If this is true, then the large
luminosity spreads we observe for { USC-near and -far} would have
little to do with an (unlikely) spread of ages in USC, and indicate
instead a wide range of past accretion histories for USC stars.
It should be remarked that the individual accretion histories in a
remote past are no
longer traceable from (for instance) accretion or disk diagnostics
observable today, and that the star positions in the CAMD need not be
altered by non-photospheric contributions for such a scatter to take place.
The large $\Delta (M_G)$ for the same stars, on the other hand, would agree
well with the younger ages of { USC and Lupus~III} among all Sco~OB2
populations, as
reported in the literature, and suggested by their compact morphology.

{
The compact groups UCL-1 and UCL-3 (and UCL-2 to a lesser
extent) fall in Fig.~\ref{age-delta-sigma} close to the diffuse
D1 and D2a populations, with which they are also nearly co-spatial
(Fig.~\ref{diffuse-space}).  This suggests that these clusters are nearly
coeval with the diffuse population in UCL, and are what remains of the
original star formation sites in that part of Sco~OB2.
}
Determining if these subclusters are
bound or not would require estimates of their masses, spatial and velocity
dispersion, which were not quantitatively made.

\section{Three-dimensional structure}
\label{three-dim}

\begin{figure*}
\resizebox{\hsize}{!}{
\includegraphics[]{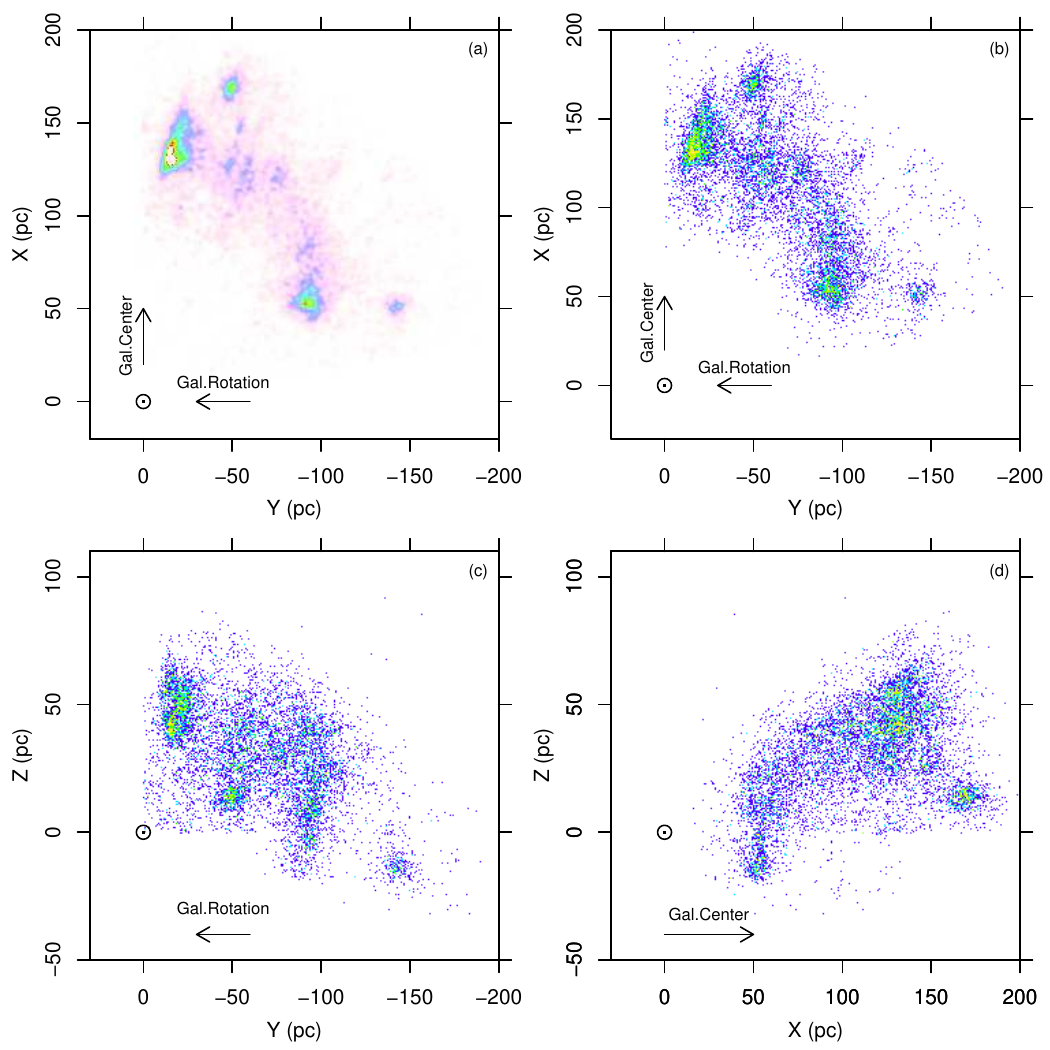}}
\caption{
Positions of all { VT-selected} PMS Sco~OB2 members
in Galactic $XYZ$ coordinates.
Panel $a$: smoothed distribution in the $XY$ plane; unlike other
panels, also upper-MS stars are included here. The Sun lies at
$(Y,X)=(0,0)$. The directions of the Galactic center and rotation are
indicated with arrows.
Panels $b,c,d$: spatial distributions of PMS members, projected onto the 
$YX$, $YZ$, and $XZ$ planes respectively, with no smoothing applied.
Density is maximum for yellow color and minimum (non-zero) for blue;
white indicates zero density.
\label{xyz-all}}
\end{figure*}

The availability of precise parallax values (to better than 10\%) for
our Sco~OB2 PMS members enables us to study their three-dimensional
(3-d) placement in space, represented using the $XYZ$ coordinate system.
The various two-dimensional projections of the full 3-d positions are
shown in Fig.~\ref{xyz-all}.
{
We show only the VT-selected members for highest reliability, from both
the PMS and Upper-MS samples in panel~$a$, and only PMS in panels~$b,c,d$.
We also included IC~2602 for comparison.
}
Panel $a$ is slightly smoothed to emphasize some details in the
higher-density regions, while the other three panels show the
un-smoothed distributions, to show even individual stars. Thanks to the
good-quality parallaxes used, only a mild finger-of-God effect
(elongated distributions towards the Sun) is
noticeable. The highest-density region in panel~$a$ coincides with the
USC region surrounding $\rho$~Oph, at $(Y,X) \sim (-15,135)$, and extends
almost 40~pc in depth.
The second-highest concentration is found at the nearest edge of LCC,
and coincides with the cluster LCC-1 at $(Y,X) \sim (-90,50)$.
This is in apparent contradiction
with the density map of Fig.~\ref{spatial-map-all-satur}, where the
second-highest density peak was at the cluster UCL-1: this is explained
by observing that UCL-1 is much farther out than LCC-1, and therefore
appears more compact on the sky while the latter appears ``diluted''; on
the contrary, shown in Fig.~\ref{xyz-all} is the true space density of
stars, irrespective of projection effects. UCL-1 (at $(Y,X) \sim
(-50,170)$) becomes thus only the third-highest density peak in real space.
USC-1 and LCC-1 lie $\sim 125$~pc apart in space, a distance which can
be considered as the total length of Sco~OB2.
The minor peaks near $(Y,X) \sim (-50,120)$ correspond to PMS stars in
the Lupus clouds.
IC~2602 is found at $(Y,X) \sim (-140,50)$: its halo is again
recognizable, and found to extend along the line of sight as it was
on the sky plane (Fig.~\ref{spatial-map-all-satur}).
{
A careful inspection of its spatial distribution reveals that its core is
double-peaked, a fact that could only be found from Gaia precise
parallaxes: the two peaks are aligned along the line of sight, and not
detected in the sky projected member distribution
(Fig.~\ref{spatial-map-all-satur}).
}

In panels~$c,d$ of the same figure,
the densest regions at $Z \sim 50$ correspond to USC. The densest cluster
at $Z<0$ is IC~2602, while UCL-1 and LCC-1 lie at $Z \sim +15$ and $\sim
0$, respectively.

\begin{figure*}
\includegraphics[width=17.5cm]{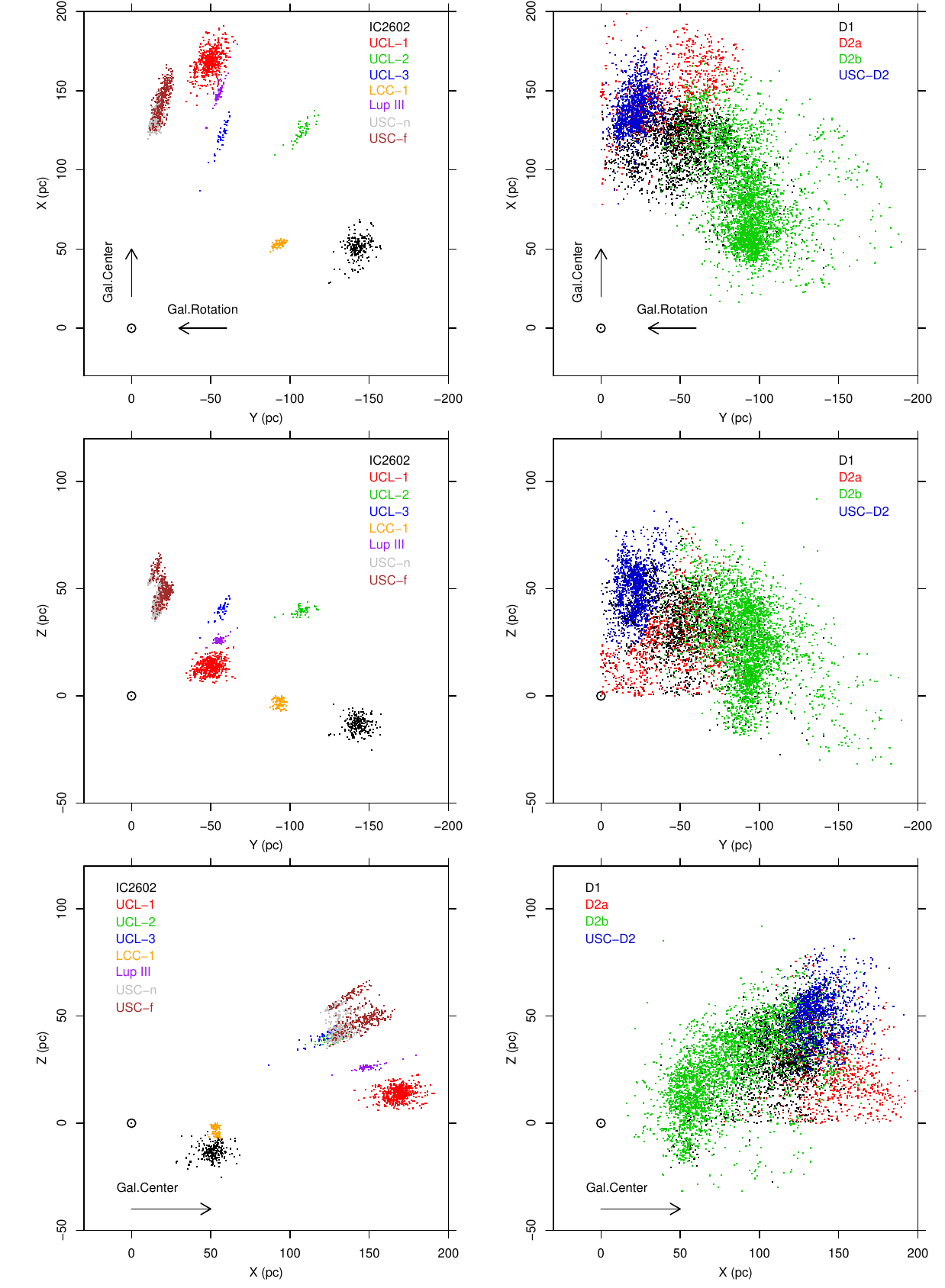}
\caption{
Left panels: projections on the $YX$, $YZ$, and $XZ$ planes of positions
of { VT-selected} PMS members in compact { populations.}
Right panels: same projections for { VT-selected} PMS stars in diffuse
{ populations.}
\label{xyz-diffuse-compact}}
\end{figure*}

As the results from the previous section indicate, the compact and diffuse
Sco~OB2 population show { both similarities and differences among
them}, and we have therefore studied their 3-d distributions separately.
This is done
in Fig.~\ref{xyz-diffuse-compact}, where panels in the left column show
the compact groups and those on the right column the diffuse groups
({ VT-selected} PMS only).
These density plots are unsmoothed to show even individual stars.
Several interesting observations may be made from these plots.
{ The IC~2602 population was selected on the basis of (loose) spatial
and PM constraints, with no selection of parallax; yet,
Fig.~\ref{xyz-diffuse-compact} shows that there are at most a handful of
stars in the IC~2602 group along its line of sight at random parallax
values. Therefore, the number of contaminants in that sample (as far as
the PMS stars are concerned) is very small.
}
The halo around
IC~2602 is therefore composed of genuine dynamical members of the
cluster, and we see from the figure that it measures up to $\sim 40$~pc
in diameter, a size never suspected for that cluster, which reinforces
our arguments above on cluster evaporation. Next,
{
UCL-1 is much more distant than IC~2602, and although it appears more
compact on the sky it has a comparable physical size from
Fig.~\ref{xyz-diffuse-compact}. Its small satellite lies at the same
distance. Again, there are virtually no contaminant field stars all
along the line of sight, despite no parallax-based selection was applied in
the cluster definition. This is true of all other compact populations as well.
The USC-near and -far groups were selected on the basis of their
different kinematics and average parallaxes, but as noted above are not
very different in terms of their spatial distribution. The two groups
are largely overlapping in physical space, a puzzling result which will
deserve a more detailed study (possibly including spectroscopic
measurements, on a larger scale than made by e.g.\ Rigliaco \e 2016).
}

The diffuse { populations} present projected spatial distributions (right
column of Fig.~\ref{xyz-diffuse-compact}) which
are { rather} unlike those of compact groups. They are on average
{ slightly} closer
to us than the compact groups { (median distance for compact groups,
excluding IC~2602: 155~pc; for diffuse populations: 143~pc).}
{
The diffuse populations, which were selected essentially based
on their 3-d kinematics, have also different distributions in space
among themselves. D1 and D2a lie at markedly different average
distances, although there is some physical mutual overlap between them.
Fig.~\ref{xyz-diffuse-compact} shows that D1 also possesses spatial
substructures, which were also suggested by our kinematical analysis
using method~B.
The USC-D2 diffuse population in USC shows the highest degrees of
spatial complexity and substructures, not an unexpected result since it
bears many similarities to the compact USC populations both in kinematics
(Sect.~\ref{diff-usc}) and CAMD (Sect.~\ref{ages}).
}

We discussed in the previous section evidence that { most of the} diffuse
populations in Sco~OB2 are older than the compact ones { in Lupus and USC}:
therefore, Fig.~\ref{xyz-diffuse-compact} also suggests that star formation in
Sco~OB2 started in regions closer to the Sun, and then continued towards
regions further away like USC, in general agreement with literature results.

{
The comparison between the left and right columns of
Fig.~\ref{xyz-diffuse-compact} shows that LCC-1 has a close connection
with the D2b population, of which it might constitute a sort of remnant
core. We examined the kinematical connection between Lup-III and the
diffuse population in Sect.~\ref{diffuse}, and found that it is only
compatible with D2b. This is also consistent with the respective space
distributions from Fig.~\ref{xyz-diffuse-compact}: in this case, Lup-III
would nearly exactly mark the D2b boundary at large $l$.
UCL-2 and -3 were found kinematically consistent with D1, but examining
their spatial distribution this association becomes problematic for
UCL-2, since D1 does not reach as far as the position of this subcluster.
}

\section{Discussion and conclusions}
\label{concl}

We have analyzed the Gaia DR2 data on a sky region of $\sim 2000$ square
degrees containing the entire Sco~OB2 association. Association members
are clearly clustered on the proper-motion
{ and transverse-velocity planes}, resolving any
ambiguity with the solar reflex motion which was present in earlier,
lower-precision astrometric data. Moreover, Gaia photometry and
precise parallaxes of initial candidate members permit a clear
distinction between true low-mass members, populating a clear PMS locus in
the Color-Absolute Magnitude Diagram, from MS field stars; no such clear
distinction is instead possible for more massive MS members of Sco~OB2.
The total number of PMS association members is of 10839 PMS stars, with
{ 1-3\%} contamination from non-members, plus 3598 upper-MS stars with a
larger contamination { (10-30\%). A less contaminated, but less
complete member sample was obtained using transverse velocities (9221 PMS and
1337 Upper-MS stars).}
Of the { PM-selected} PMS members,
2862 fall inside the
conventional boundaries of Upper Sco-Cen sub-association, 4511 in the
Upper Centaurus-Lupus, and 2803 in Lower Centaurus-Crux.
This is the largest and most complete population ever found in Sco~OB2,
down to the bottom of the stellar mass spectrum or even below.
Most of the member stars are found in a spatially diffuse component, on
which local density enhancements are superimposed.
The association spans a large arc, both projected on the sky and in
actual space, in good agreement with earlier works (e.g., Sartori \e
2003), but with much better detail, and member distances range from
approximately 100~pc to almost 200~pc.

The proper motion and parallax distributions of members are clearly
resolved thanks to the small measurement errors, and show considerable
substructures. We have identified
{
7 subclusters (plus IC~2602) which are compact both in physical and
proper-motion space. We have also applied a new method to reconstruct the 3-d
kinematics of diffuse populations, finding 4-5 distinct populations with
distinct spatial and kinematical properties. These do not correspond to
the traditional boundaries between USC, UCL and LCC
sub-associations, their spatial distribution being more complex.
}
The most
``kinematically'' compact group is a cluster near V1062~Sco, recently
discovered by R\"oser \e (2018), and here named UCL-1: still, this
cluster { was resolved spatially (and in part, kinematically),}
with a probable satellite sub-cluster
only a few degrees away. While R\"oser \e (2018) list 63 stars from Gaia
DR1 data, we find a total of 593 { PMS$+$Upper-MS members in UCL-1.}
The densest, but not richest part of Sco~OB2 is
Upper Sco, composed of an incoherent (both in proper-motion space and in
real space) assembly of local stellar groupings, lacking a well-defined
global organization.

We find evidence from the CAMD
{
that the majority (3/4) of the
diffuse populations are older than compact populations in USC and Lupus:
}
this fits with the picture of stars diffusing away rapidly from
their formation sites. We find however a large spread in apparent ages
from the CAMDs of { the youngest} groups, as if the same
compact cloud had formed stars during a long period of time: this fits
less well together with the age ordering based on kinematics. A possible
solution that we favor is that the luminosity spread in the PMS region
of the CAMD has its origin in the past accretion histories of individual
stars, in agreement with recent theoretical models.

{
The distribution of PMS stars on the CAMD is consistent with
the sequence of star-formation events, from LCC and UCL
to USC, reported in the literature.
We find no significant age difference between the diffuse
populations in LCC and UCL, in agreement with Mamajek \e (2002) and PM16.
}

In our study we included, almost serendipitously, the PMS stars in the
open cluster IC~2602, near the western edge of Sco~OB2 and close to it
on the PM plane: the parent clouds of Sco~OB2 and IC~2602 might be
considered as close relatives. We find that the cluster has a double
core, with two distinct parallax peaks; moreover, it possess an extended
low-density halo ($>10^{\circ}$ in total size), which was never reported
before to our knowledge: its detection was possible only due to the
unique capabilities of Gaia.

Last, Gaia enables us to determine with accuracy the number of field
stars, down to the same mass limits, which are co-spatial with Sco~OB2
members, not only in sky projection but in actual three-dimensional
space. For example, we find { more than 2500} MS stars in the USC region
between $345<l<355$, $12<b<25$, and $6<\pi<8$~mas, where { only 1977}
PMS members are found. Even worse, within the UCL sky limits and
parallax limits $6.5<\pi<10$~mas, more than 21000 MS field
stars outnumber the 2600 PMS UCL members { in the same parallax
range}. This means that very early
during the evolution of these stars their dynamics will be dominated by
the general Galactic potential, and not by the gravity of their coeval
association members. It is interesting, however, that despite being so
diffuse on the sky and spatially mixed with field stars, Sco~OB2 members
still keep a strong memory of their initial kinematics, which will be
gradually erased by dynamical friction. These data would permit an
observational test of theoretical N-body models of mixing two
co-spatial, but kinematically distinct star populations (e.g., Mapelli
\e 2015).

The multi-peaked structure in the PM plane of stars in USC
{ (Fig.~\ref{usc-compact-pm}), and their highly asymmetric spatial
distribution (Fig.~\ref{usc-compact-space}),}
suggest that turbulence and irregular geometries play a major role in
determining the dynamical properties of the newly formed stars in USC, while
global ordered motions are much less important.
Considering an approximate USC size of 10 degrees, and a $\mu_b$ dispersion
$\sim 1.4$~mas/yr, the crossing time is $t_{\rm cross} \sim 25$~Myr, and
using the formula $t_{\rm relax} = \frac{N}{8 \ln (N)} t_{\rm cross}$
(e.g., Binney and Tremaine 1987)
we obtain a relaxation time scale $t_{\rm relax} \sim 1200$~Myr, using
$N=2862$ from { Table~\ref{table-breakdown}} as the number of USC members:
it is likely that USC stars will
disperse before relaxation is attained, because of evaporation and
release of binding energy during gas dispersal.

\begin{acknowledgements}
An anonymous referee made challenging yet stimulating comments which
resulted in substantial improvements in the paper.
This work presents results from the European Space Agency (ESA) space
mission Gaia. Gaia data are being processed by the Gaia Data Processing
and Analysis Consortium (DPAC). Funding for the DPAC is provided by
national institutions, in particular the institutions participating
in the Gaia MultiLateral Agreement (MLA). The Gaia mission website
is https://www.cosmos.esa.int/gaia. The Gaia archive website is
https://archives.esac.esa.int/gaia.
This research makes use of the SIMBAD database and the Vizier catalog service,
operated at CDS, Strasbourg, France.
We also make heavy use of R: A language and environment for statistical
computing. R Foundation for Statistical Computing, Vienna, Austria.
(http://www.R-project.org/).
\end{acknowledgements}

\bibliographystyle{aa}

\begin{thebibliography}{}

\bibitem[Arenou et al.(2018)]{2018arXiv180409375A} Arenou, F., Luri, X., Babusiaux, C., et al.\ 2018, arXiv:1804.09375

\bibitem[Baraffe et al.(2012)]{2012ApJ...756..118B} Baraffe, I., Vorobyov, E., \& Chabrier, G.\ 2012, \apj, 756, 118

\bibitem[Baraffe et al.(2017)]{2017A&A...597A..19B} Baraffe, I., Elbakyan, V.~G., Vorobyov, E.~I., \& Chabrier, G.\ 2017, \aap, 597, A19

\bibitem[Binney \& Tremaine(1987)]{1987gady.book.....B} Binney, J., \&
Tremaine, S.\ 1987, Princeton, NJ, Princeton University Press, 1987

\bibitem[Blaauw(1964)]{1964ARA&A...2..213B} Blaauw, A.\ 1964, \araa, 2, 213


\bibitem[Bravi et al.(2018)]{2018A&A...615A..37B} Bravi, L., Zari, E., Sacco, G.~G., et al.\ 2018, \aap, 615, A37

\bibitem[Chambers et al.(2016)]{2016arXiv161205560C} Chambers, K.~C.,
Magnier, E.~A., Metcalfe, N., et al.\ 2016, arXiv:1612.05560


\bibitem[Damiani et al.(2017)]{2017A&A...602L...1D} Damiani, F., Prisinzano, L., Jeffries, R.~D., et al.\ 2017, \aap, 602, L1

\bibitem[de Bruijne(1999)]{1999MNRAS.310..585D} de Bruijne, J.~H.~J.\ 1999, \mnras, 310, 585

\bibitem[Dehnen \& Binney(1998)]{1998MNRAS.298..387D} Dehnen, W., \& Binney, J.~J.\ 1998, \mnras, 298, 387

\bibitem[de Zeeuw et al.(1999)]{1999AJ....117..354D} de Zeeuw, P.~T., Hoogerwerf, R., de Bruijne, J.~H.~J., Brown, A.~G.~A., \& Blaauw, A.\ 1999, \aj, 117, 354

\bibitem[Dobbie et al.(2010)]{2010MNRAS.409.1002D} Dobbie, P.~D., Lodieu, N., \& Sharp, R.~G.\ 2010, \mnras, 409, 1002

\bibitem[Drew et al.(2014)]{2014MNRAS.440.2036D} Drew, J.~E.,
Gonzalez-Solares, E., Greimel, R., et al.\ 2014, \mnras, 440, 2036

\bibitem[Dunham \& Vorobyov(2012)]{2012ApJ...747...52D} Dunham, M.~M., \& Vorobyov, E.~I.\ 2012, \apj, 747, 52

\bibitem[Famaey et al.(2005)]{2005A&A...430..165F} Famaey, B., Jorissen, A., Luri, X., et al.\ 2005, \aap, 430, 165

\bibitem[Gaia Collaboration et al.(2016)]{2016A&A...595A...1G} Gaia
Collaboration, Prusti, T., de Bruijne, J.~H.~J., et al.\ 2016, \aap, 595, A1

\bibitem[Gaia Collaboration et al.(2018)]{2018arXiv180409365G} Gaia
Collaboration, Brown, A.~G.~A., Vallenari, A., et al.\ 2018a,
arXiv:1804.09365

\bibitem[Gaia Collaboration et al.(2018)]{2018arXiv180409382G} Gaia Collaboration, Eyer, L., Rimoldini, L., et al.\ 2018b, arXiv:1804.09382

\bibitem[Goldman et al.(2018)]{2018arXiv180702076G} Goldman, B., Roeser, S., Schilbach, E., Moor, A.~C., \& Henning, T.\ 2018, arXiv:1807.02076

\bibitem[Hoogerwerf \& Aguilar(1999)]{1999MNRAS.306..394H} Hoogerwerf, R., \& Aguilar, L.~A.\ 1999, \mnras, 306, 394

\bibitem[Hoogerwerf(2000)]{2000MNRAS.313...43H} Hoogerwerf, R.\ 2000, \mnras, 313, 43

\bibitem[Jeffries(2012)]{2012ASSP...29..163J} Jeffries, R.~D.\ 2012, Astrophysics and Space Science Proceedings, 29, 163

\bibitem[Jeffries et al.(2014)]{2014A&A...563A..94J} Jeffries, R.~D., Jackson, R.~J., Cottaar, M., et al.\ 2014, \aap, 563, A94

\bibitem[Johnson \& Soderblom(1987)]{1987AJ.....93..864J} Johnson, D.~R.~H., \& Soderblom, D.~R.\ 1987, \aj, 93, 864

\bibitem[Kharchenko et al.(2013)]{2013A&A...558A..53K} Kharchenko, N.~V., Piskunov, A.~E., Schilbach, E., R{\"o}ser, S., \& Scholz, R.-D.\ 2013, \aap, 558, A53

\bibitem[Kounkel et al.(2018)]{2018arXiv180504649K} Kounkel, M., Covey, K., Su{\'a}rez, G., et al.\ 2018, arXiv:1805.04649

\bibitem[Krautter et al.(1997)]{1997A&AS..123..329K} Krautter, J., Wichmann, R., Schmitt, J.~H.~M.~M., et al.\ 1997, \aaps, 123, 329

\bibitem[Lada \& Lada(2003)]{2003ARA&A..41...57L} Lada, C.~J., \& Lada, E.~A.\ 2003, \araa, 41, 57

\bibitem[Mamajek et al.(2002)]{2002AJ....124.1670M} Mamajek, E.~E., Meyer, M.~R., \& Liebert, J.\ 2002, \aj, 124, 1670

\bibitem[Manara et al.(2018)]{2018A&A...615L...1M} Manara, C.~F., Prusti, T., Comeron, F., et al.\ 2018, \aap, 615, L1

\bibitem[Mapelli et al.(2015)]{2015A&A...578A..35M} Mapelli, M., Vallenari, A., Jeffries, R.~D., et al.\ 2015, \aap, 578, A35

\bibitem[Pecaut et al.(2012)]{2012ApJ...746..154P} Pecaut, M.~J., Mamajek, E.~E., \& Bubar, E.~J.\ 2012, \apj, 746, 154

\bibitem[Pecaut \& Mamajek(2016)]{2016MNRAS.461..794P} Pecaut, M.~J., \& Mamajek, E.~E.\ 2016, \mnras, 461, 794

\bibitem[Pillitteri et al.(2016)]{2016A&A...592A..88P} Pillitteri, I.,
Wolk, S.~J., Chen, H.~H., \& Goodman, A.\ 2016, \aap, 592, A88

\bibitem[Preibisch \& Mamajek(2008)]{2008hsf2.book..235P} Preibisch, T., \& Mamajek, E.\ 2008, Handbook of Star Forming Regions, Volume II, 5, 235

\bibitem[Rigliaco et al.(2016)]{2016A&A...588A.123R} Rigliaco, E., Wilking, B., Meyer, M.~R., et al.\ 2016, \aap, 588, A123

\bibitem[Rizzuto et al.(2011)]{2011MNRAS.416.3108R} Rizzuto, A.~C., Ireland, M.~J., \& Robertson, J.~G.\ 2011, \mnras, 416, 3108

\bibitem[R{\"o}ser et al.(2018)]{2018A&A...614A..81R} R{\"o}ser, S., Schilbach, E., Goldman, B., et al.\ 2018, \aap, 614, A81

\bibitem[Sacco et al.(2015)]{2015A&A...574L...7S} Sacco, G.~G., Jeffries, R.~D., Randich, S., et al.\ 2015, \aap, 574, L7

\bibitem[Sartori et al.(2003)]{2003A&A...404..913S} Sartori, M.~J., L{\'e}pine, J.~R.~D., \& Dias, W.~S.\ 2003, \aap, 404, 913

\bibitem[Sciortino et al.(1998)]{1998A&A...332..825S} Sciortino, S., Damiani, F., Favata, F., \& Micela, G.\ 1998, \aap, 332, 825

\bibitem[Tobin et al.(2015)]{2015AJ....149..119T} Tobin, J.~J., Hartmann, L., F{\H u}r{\'e}sz, G., Hsu, W.-H., \& Mateo, M.\ 2015, \aj, 149, 119

\bibitem[van Leeuwen(2009)]{2009A&A...497..209V} van Leeuwen, F.\ 2009, \aap, 497, 209

\bibitem[Wright \& Mamajek(2018)]{2018MNRAS.476..381W} Wright, N.~J., \& Mamajek, E.~E.\ 2018, \mnras, 476, 381

\end{thebibliography}

\begin{landscape}
\begin{table}
\caption{Gaia data for PMS members of Sco~OB2.
Column {\it Sel} has {\it p} for PM-selection, and {\it pv} for PM- and
VT-selection. Column {\it Pop} indicates the association sub-population.
Full table in electronic format only.}
\label{table-pms}
\begin{tabular}{rccccccccccccl}
  \hline
Seq & Designation & RA & Dec & l & b & Parallax & $\mu_l$ & $\mu_b$ & $G$ & $BP-RP$ & Sel.\ & Pop.\ & SIMBAD \\
no.\ & Gaia DR2 & (J2000) & (J2000) & & & (mas) & (mas/yr) & (mas/yr) & (mag) & (mag) & & & Ident.\ \\
  \hline
  1 & 5299358596621466752 & 140.86254 & -60.64461 & 280.00657 & -7.40057 & 7.34 & -23.05 & -1.99 & 15.97 & 3.24 & p &  &  \\ 
    2 & 5364724700323400704 & 157.66940 & -48.15017 & 280.06253 & 8.37233 & 7.63 & -23.31 & -11.75 & 14.48 & 2.47 & pv & D2b &  \\ 
    3 & 5364348044571002880 & 157.60177 & -48.30253 & 280.10348 & 8.21870 & 7.55 & -23.21 & -11.50 & 16.69 & 3.60 & pv & D2b &  \\ 
    4 & 5298959851855997440 & 140.45557 & -61.02387 & 280.13702 & -7.80855 & 6.75 & -18.59 & -3.26 & 16.29 & 3.16 & p &  &  \\ 
    5 & 5305964428135613696 & 146.33522 & -57.80278 & 280.14232 & -3.41561 & 6.17 & -16.83 & 0.53 & 15.42 & 2.75 & p &  &  \\ 
    6 & 5298936968269001088 & 139.72309 & -61.38697 & 280.15061 & -8.31464 & 7.30 & -22.33 & -1.17 & 14.63 & 2.62 & p &  &  \\ 
    7 & 5299342550623809536 & 141.10164 & -60.73504 & 280.15454 & -7.38197 & 7.36 & -23.40 & -2.36 & 16.80 & 3.23 & p &  &  \\ 
    8 & 5297335289063109376 & 137.85486 & -62.35850 & 280.25841 & -9.62154 & 6.71 & -18.70 & -1.68 & 16.61 & 3.42 & p &  &  \\ 
    9 & 5259622040234792960 & 149.06699 & -56.19517 & 280.28069 & -1.22985 & 6.34 & -16.49 & -1.38 & 16.55 & 3.13 & p &  &  \\ 
   10 & 5356723485508902528 & 153.77877 & -52.48131 & 280.31257 & 3.38716 & 5.59 & -31.38 & -14.52 & 14.19 & 2.18 & p &  &  \\ 
   11 & 5298797227209724928 & 138.43246 & -62.21269 & 280.33373 & -9.32506 & 7.48 & -21.95 & -1.01 & 17.52 & 3.56 & p &  &  \\ 
   12 & 5356728158433450880 & 154.00672 & -52.35504 & 280.35697 & 3.56967 & 6.62 & -17.89 & -3.26 & 15.33 & 2.75 & p &  &  \\ 
   13 & 5356728158433450624 & 154.00732 & -52.35575 & 280.35767 & 3.56928 & 6.66 & -18.23 & -3.72 & 15.36 & 2.79 & p &  &  \\ 
   14 & 5259585515831918208 & 148.61511 & -56.64572 & 280.36308 & -1.73844 & 5.83 & -15.51 & -1.15 & 16.41 & 3.15 & p &  &  \\ 
   15 & 5297323881630045312 & 137.43431 & -62.69393 & 280.37887 & -9.99041 & 7.10 & -37.80 & -6.98 & 16.04 & 3.02 & p &  &  \\ 
   16 & 5305809362631987968 & 145.51876 & -58.71937 & 280.41452 & -4.39071 & 7.15 & -20.17 & -5.38 & 13.62 & 2.17 & p &  &  \\ 
   17 & 5299141202558870784 & 140.94130 & -61.18180 & 280.41707 & -7.75335 & 6.18 & -23.34 & 3.25 & 17.26 & 3.25 & p & IC2602 &  \\ 
   18 & 5364161063178792320 & 156.79744 & -49.78068 & 280.43040 & 6.68320 & 6.46 & -40.56 & -13.62 & 15.54 & 2.78 & p &  &  \\ 
   19 & 5358221260857329024 & 154.74296 & -51.86578 & 280.46170 & 4.22761 & 5.34 & -45.49 & -17.61 & 15.35 & 2.52 & p &  &  \\ 
   20 & 5299147799628892928 & 141.48328 & -60.99549 & 280.47243 & -7.43656 & 7.04 & -21.79 & -3.11 & 14.20 & 2.50 & p &  &  \\ 
   \hline
\end{tabular}
\end{table}
\end{landscape}

\begin{landscape}
\begin{table}
\caption{Gaia data for Upper-MS members of Sco~OB2.
Column {\it Sel} has {\it p} for PM-selection, and {\it pv} for PM- and
VT-selection. Column {\it Pop} indicates the association sub-population.
Full table in electronic format only.}
\label{table-upp-ms}
\begin{tabular}{rccccccccccccl}
  \hline
Seq & Designation & RA & Dec & l & b & Parallax & $\mu_l$ & $\mu_b$ & $G$ & $BP-RP$ & Sel.\ & Pop.\ & SIMBAD \\
no.\ & Gaia DR2 & (J2000) & (J2000) & & & (mas) & (mas/yr) & (mas/yr) & (mag) & (mag) & & & Ident.\ \\
  \hline
10840 & 5297338690676935424 & 137.45916 & -62.28603 & 280.07878 & -9.70917 & 5.08 & -21.14 & -11.75 & 9.25 & 0.45 & p &  & HD 79208 \\ 
  10841 & 3482758885492572288 & 169.83631 & -29.75181 & 280.08724 & 29.00249 & 5.39 & -45.92 & -14.03 & 10.96 & 0.75 & p &  & TYC 6662-147-1 \\ 
  10842 & 5400254765733986560 & 166.06918 & -36.72145 & 280.10706 & 21.35562 & 7.06 & -40.41 & -7.97 & 7.05 & 0.33 & p &  & HD 96054 \\ 
  10843 & 5356354633714239232 & 151.72070 & -53.91337 & 280.12459 & 1.50700 & 9.22 & -40.98 & -10.19 & 7.91 & 0.58 & pv & D1 & HD 87933 \\ 
  10844 & 5297334773667021184 & 137.71401 & -62.34676 & 280.20501 & -9.66202 & 8.94 & -41.83 & -11.53 & 11.10 & 1.27 & pv & D1 & TYC 8944-1821-1 \\ 
  10845 & 5297336148056458496 & 137.81932 & -62.31691 & 280.21595 & -9.60575 & 6.22 & -36.90 & -17.16 & 3.88 & -0.24 & p &  &  i Car \\ 
  10846 & 5364840733158545792 & 158.96053 & -47.09885 & 280.27595 & 9.72067 & 8.27 & -24.06 & -16.07 & 7.92 & 0.55 & p &  & HD 91922 \\ 
  10847 & 5391528556121578880 & 162.83895 & -42.11241 & 280.28803 & 15.41879 & 6.00 & -14.89 & -11.65 & 12.17 & 1.13 & p &  & UCAC2 13751697 \\ 
  10848 & 5299184358391332096 & 141.66186 & -60.67638 & 280.30984 & -7.14778 & 5.04 & -22.37 & 2.00 & 9.22 & 0.58 & p & IC2602 & HD 81991 \\ 
  10849 & 5305312653936008704 & 143.24813 & -59.92524 & 280.36018 & -6.06185 & 6.16 & -36.96 & -5.19 & 12.06 & 1.17 & p &  &  \\ 
  10850 & 5393107630917045888 & 163.75225 & -41.01782 & 280.40006 & 16.70466 & 5.28 & -40.30 & -6.37 & 11.62 & 0.91 & p &  & TYC 7728-1101-1 \\ 
  10851 & 5299141546145254528 & 140.89550 & -61.19323 & 280.40948 & -7.77704 & 10.31 & -31.47 & -7.10 & 9.73 & 1.05 & p &  &  V479 Car \\ 
  10852 & 5358156252226252672 & 156.78337 & -49.78759 & 280.42631 & 6.67254 & 6.32 & -41.98 & -14.88 & 8.28 & 0.24 & p &  & HD 90692 \\ 
  10853 & 5390102287680538112 & 163.69873 & -41.17828 & 280.43824 & 16.54330 & 7.09 & -46.61 & -11.63 & 10.50 & 0.82 & p &  & TYC 7728-1105-1 \\ 
  10854 & 5298784548466453376 & 138.30205 & -62.42212 & 280.44796 & -9.51170 & 6.20 & -22.73 & 3.54 & 9.05 & 0.63 & p & IC2602 & HD 79796 \\ 
  10855 & 5298784479746978816 & 138.33119 & -62.42979 & 280.46295 & -9.50699 & 6.21 & -23.41 & 3.80 & 10.31 & 0.73 & p &  & HD 309549 \\ 
  10856 & 5391005124166906240 & 162.18278 & -43.42229 & 280.47680 & 14.03515 & 5.59 & -16.41 & -13.82 & 8.70 & 0.41 & p &  & HD 93749 \\ 
  10857 & 5356043368830386816 & 151.78360 & -54.56351 & 280.53501 & 1.00137 & 5.30 & -30.34 & -11.27 & 9.71 & 0.61 & p &  &  \\ 
  10858 & 5356043368839496704 & 151.78442 & -54.56375 & 280.53554 & 1.00145 & 5.21 & -22.98 & -7.32 & 7.68 & 0.16 & pv & D2b &  \\ 
  10859 & 5404015542177210880 & 169.72006 & -31.02597 & 280.55713 & 27.79307 & 5.59 & -25.81 & -11.75 & 12.24 & 1.25 & p &  &  \\ 
   \hline
\end{tabular}
\end{table}
\end{landscape}

\end{document}